\documentclass{cimento}
\newcommand{\apjl}{ApJ}
\newcommand{\apjs}{ApJS}
\newcommand{\apj}{ApJ}
\newcommand{\mnras}{MNRAS}
\newcommand{\aap}{A\&A} 
\newcommand{\jcap}{JCAP}
\newcommand{\pasj}{PASJ} 
\newcommand{\physrep}{Phys. Rep.} 
\newcommand{\araa}{Ann. Rev. Astron. Astrophys.}

\newcommand{\simgt}{\lower.5ex\hbox{$\; \buildrel > \over \sim \;$}}
\newcommand{\simlt}{\lower.5ex\hbox{$\; \buildrel < \over \sim \;$}}

 \newcommand{\Real}[1]{{\rm Re}\left[ #1 \right]}

 \newcommand{\baredth}{\;\overline{\raise1.0pt\hbox{$'$}\hskip-6pt \partial}\;}
 \newcommand{\edth}{\;\raise1.0pt\hbox{$'$}\hskip-6pt\partial\;}
 \newcommand{\bx}{\mbox{\boldmath $x$}}
 \newcommand{\bk}{\mbox{\boldmath $k$}}
 
 \newcommand{\bchi}{\mbox{\boldmath $\chi$}}
 \newcommand{\bbeta}{\mbox{\boldmath $\beta$}}
 \newcommand{\btheta}{\mbox{\boldmath $\theta$}}
 \newcommand{\bnabla}{\mbox{\boldmath $\nabla$}}  
 \newcommand{\balpha}{\mbox{\boldmath $\alpha$}}
 \newcommand{\boldeta}{\mbox{\boldmath $\eta$}}  
 \newcommand{\bxi}{\mbox{\boldmath $\xi$}}
 \newcommand{\bA}{\mbox{\boldmath ${A}$}}
 \newcommand{\bI}{\mbox{\boldmath ${I}$}}
 \newcommand{\bH}{\mbox{\boldmath ${H}$}}
 
 \newcommand{\bPsi}{\mbox{\boldmath ${\Psi}$}}
 
\usepackage{graphicx}  
\title{Cluster Weak Gravitational Lensing}
\author{Keiichi~Umetsu\from{ins:x}\from{ins:y}
}
\instlist{\inst{ins:x} Academia Sinica Institute of Astronomy and
Astrophysics, P.O. Box 23-141, Taipei 10617, Taiwan
  \inst{ins:y} Leung center for Cosmology and Particle Astrophysics,
  National Taiwan University, Taipei 10617, Taiwan} 
\begin{document}

\maketitle

\begin{abstract}
Weak gravitational lensing of background galaxies 
is a unique, direct probe of the distribution of matter
in clusters of galaxies.
We review several important aspects 
of cluster weak gravitational lensing together with
recent advances in weak lensing techniques for measuring cluster lensing
profiles and constraining cluster structure parameters.
\end{abstract}
\tableofcontents

\newpage

\section{Introduction}

Propagation of light rays from a distant source to the observer is
governed by the gravitational field of intervening mass fluctuations
as well as by the global geometry of the universe.
The images of background sources 
hence carry the imprint of the gravitational potential 
of intervening cosmic structures, and
their statistical properties
can be used to test the background cosmological models.

The deep gravitational potential wells of clusters of galaxies
generate weak shape distortions of the images of background
sources due to differential deflection of light rays, resulting in a
systematic distortion pattern of background source images around the
center of massive clusters, known as weak gravitational lensing\cite{1990ApJ...349L...1T,1993ApJ...404..441K,1995A&A...294..411S,2001PhR...340..291B,1999PThPS.133...53U}.
In the past decade, weak lensing has become a powerful,
reliable measure to map the distribution of matter in clusters,
dominated by invisible dark matter (DM),
without requiring any assumption about the physical and
dynamical state of
the system\cite{Clowe+2006_Bullet,Okabe&Umetsu08}.
Recently, cluster weak
lensing has been used to examine the form of DM density profiles\cite{BTU+05,UB2008,2007ApJ...668..643L,2008arXiv0805.2552M,BUM+08,2009arXiv0903.1103O,Oguri+2009_Subaru,2007arXiv0709.1159J},
aiming for an observational test of the equilibrium density profile of
DM halos and the scaling relation between halo mass and concentration, 
predicted by $N$-body simulations in the standard Lambda Cold Dark
Matter ($\Lambda$CDM) model\cite{2007ApJS..170..377S,2008arXiv0803.0547K}. 
Weak lensing techniques have also been used to search for cluster-sized
mass concentrations projected 
on the sky\cite{1996MNRAS.283..837S,2000A&A...355...23E,2000ApJ...539L...5U},   
allowing us to define samples of
``shear-selected'' DM halos from deep optical
surveys\cite{2002ApJ...580L..97M,2007ApJ...669..714M,Hamana+2009}. 

In this lecture we briefly review several important aspects 
of cluster weak gravitational lensing.
There have been several reviews of relevant subjects:
For general treatments of gravitational lensing, we refer the reader to
Schneider, Ehlers, \& Falco\cite{SEF1992}, 
Blandford \& Narayan\cite{1992ARA&A..30..311B}, 
Refsdal \& Surdej\cite{refsdal_surdej}, 
and
Narayan \& Bartelmann\cite{1996astro.ph..6001N}.
For a review on strong gravitational lensing in clusters, 
see Hattori, Kneib, \& Makino\cite{1999PThPS.133....1H}.
For a general review of weak gravitational lensing, see
Bartelmann \& Schneider\cite{2001PhR...340..291B}.

\section{Basic Equations of Cosmological Gravitational Lensing}

\begin{figure}[htb]
\begin{center}
\includegraphics[width=100mm]{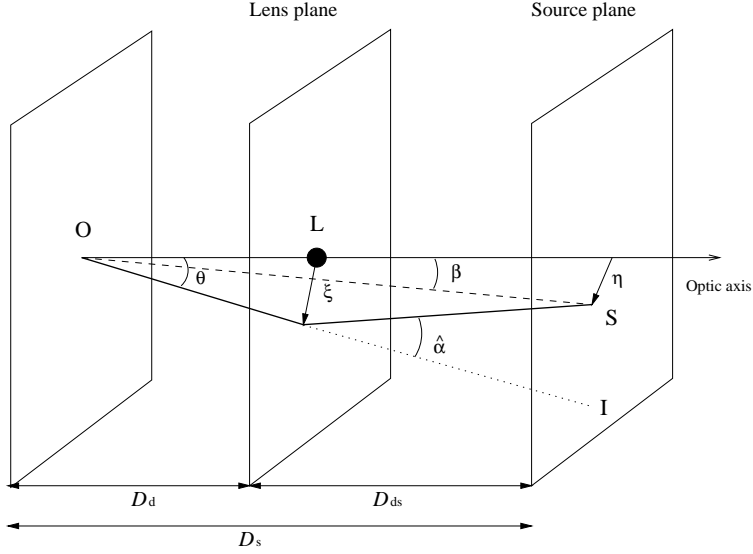}
\end{center} 
\caption{
\label{fig:lens} 
Illustration of a gravitational lensing system. The light ray
 propagates from the source {\sl S} 
at the position $\boldeta$ in the source
 plane  to the observer {\sl O}, 
passing the position $\bxi$ in the lens plane  
where it is deflected by an angle $\hat{\balpha}$. 
The angular position of the source {\sl S} relative to  the optic 
 axis is denoted by $\bbeta$, and
 that of the image {\sl I} relative to the optical axis is denoted by
 $\btheta$.  The
 angular diameter distances between the observer and lens, between the
 lens and source, and between the observer and source are $D_{d}$,
 $D_{ds}$, and $D_{s}$, respectively.}
\end{figure} 

The local universe appears to be highly inhomogeneous on a wide range
of scales from stars, galaxies, through galaxy clusters, up to
forming superclusters, filaments, and low-density voids.
The propagation of a light ray is therefore
influenced by the gravitational field caused by these inhomogeneous mass
distributions. The light propagation in an arbitrary curved space time
is in general a complicated theoretical problem. However, most of
the astrophysical relevant situations allow for a much simpler description of
the light ray propagation, which is called the gravitational lensing
theory. This section reviews briefly the gravitational lensing theory in
order to provide a foundation for later discussions on cluster weak
lensing as well as to summarize the basic equations and concepts on
cosmological gravitational lensing.

\subsection{Gravitational Deflection and Lens Equation}   
\label{sec:gl_sw} 
   
To begin with, let us consider 
the bending of light rays in the asymptotically flat spacetime
caused by 
a quasi-stationary/spatially-isolated mass distribution, assuming that
the gravitational field is weak or the deflection angle is small.
This can be done by solving the null geodesic equation in the Minkowski
spacetime perturbed with the Newtonian gravitational potential  
(see, e.g., Ref.~\cite{SEF1992}).
The perturbed metric $g_{\mu\nu}$ ($\mu,\nu=0,1,..,3$)
is then written as
\begin{equation}
\label{eq:metric_sw} 
ds^2= g_{\mu\nu} dx^\mu dx^\nu=
-(1+2\Psi/c^2)c^2 dt^2 
+(1+2\Psi/c^2)^{-1}\delta_{ij} dx^idx^j, 
\end{equation}
where 
$\Psi$  
is the Newtonian gravitational potential, 
$\delta_{ij}$ ($i,j=1,2,3$) is the Kronecker delta,
and $c$ is the speed of light.
We consider the metric given by equation (\ref{eq:metric_sw}) 
to be the sum of a
background metric $g_{\mu\nu}^{(b)}$ and a small perturbation
$h_{\mu\nu}$: $g_{\mu\nu}=g_{\mu\nu}^{(b)} + h_{\mu\nu}$ with
$|h_{\mu\nu}|\ll 1$. 
Expanding to the first order in $\Psi/c^2$ we break
up the metric in equation (\ref{eq:metric_sw}) into a sum of the background
Minkowski metric $g_{\mu\nu}^{(b)}=
\eta_{\mu\nu}={\rm diag}(-1,1,1,1)$ and a perturbation of the
form: $h_{\mu\nu}={\rm
diag}(-2\Psi/c^2,-2\Psi/c^2,-2\Psi/c^2,-2\Psi/c^2)$.
As usual,
$g^{\mu\nu}$ and $g^{(b)\mu\nu}$ are defined by $g^{\mu\rho}
g_{\rho\nu}= \delta^{\mu}_{\nu}$ and
$g^{(b)\mu\rho}
g^{(b)}_{\rho\nu}= \delta^{\mu}_{\nu}$.
Then, to the first order of $h$,
$g^{\mu\nu}= g^{(b)\mu\nu}-h^{\mu\nu}$, where $h^{\mu\nu}$ is defined by
$h^{\mu\nu}= g^{(b)\mu\rho} g^{(b)\nu\sigma} h_{\rho\sigma}$.

The propagation of light rays is described by the null geodesic equation:
\begin{eqnarray}
\label{eq:4mom}
k^\mu &\equiv& \frac{dx^\mu(\lambda)}{d\lambda},\\
\label{eq:null}
0&=&g_{\mu\nu}k^\mu k^\nu,\\
\label{eq:geodesic}
\frac{dk^\mu}{d\lambda} &=&
 -\Gamma^{\mu}_{\nu\lambda}k^\nu k^\lambda,
\end{eqnarray}
where $k^\mu$ is the 4-momentum of the light ray,
$\lambda$ is the affine parameter, and
$\Gamma^\mu_{\nu\lambda}$ denotes the Christoffel symbol:
$\Gamma^{\mu}_{\nu\rho} = (1/2) g^{\mu\lambda}
\left(
g_{\lambda \nu,\rho} + g_{\lambda \rho, \nu} - g_{\nu\rho,\lambda}
\right)$.
In the background Minkowski spacetime with
$g_{\mu\nu}^{(b)}=\eta_{\mu\nu}$,  $\Gamma^{(b)\mu}_{\nu\rho}=0$.
For a light ray moving into the positive $x$-direction
in the background spacetime,
the photon 4 momentum $k^{(b)\mu}$ and the
corresponding unperturbed orbit $x^{(b)\mu}=(ct,\bx)$ 
are simply given as
$k^{(b)\mu}=dx^{(b)\mu}/d\lambda
=(1,1,0,0)$ and $x^{(b)\mu}=(\lambda,\lambda,0,0)$.

Now we consider the light ray propagation in the perturbed space time.
Without loss of generality, we can take the deflection angle to
lie in the $xz$ plane (i.e., $y\equiv x^2=0$).
In the weak field limit ($|\Psi/c^2|\ll 1$) 
the impact parameter $b$ of the incoming light ray 
is much greater than the Schwarzschild radius of the deflector, 
$b\gg 2GM/c^2$.
The linearized null geodesic equation takes the form:
\begin{eqnarray}
\label{eq:4mom_pert}
k^\mu(\lambda) &=& k^{(b)\mu}(\lambda) + \delta k^\mu(\lambda),\\
\label{eq:null_pert}
0&=&h_{\mu\nu}k^{(b)\mu} k^{(b)\nu}+ 2g^{(b)}_{\mu\nu}
k^{(b)\mu}\delta k^\nu,\\ 
\label{eq:geodesic_pert}
\frac{d(\delta k^\mu)}{d\lambda} &=&
 -2\Gamma^{(b)\mu}_{\nu\lambda} k^{(b)\nu} \delta k^\lambda
-\delta \Gamma^{\mu}_{\nu\lambda} k^{(b)\nu} k^{(b)\lambda}.
\end{eqnarray}
Note that the first term on the right-hand side of equation
(\ref{eq:geodesic_pert}) vanishes since $\Gamma^{(b)}=0$.
The perturbed Christoffel symbol is
$\delta  \Gamma^{\mu}_{\nu\rho}=
(1/2)
\eta^{\mu\lambda}\left(
h_{\lambda \nu,\rho} +
h_{\lambda \rho, \nu}-
h_{\nu\rho,\lambda}
\right) + O(h^2)$.
Choosing the boundary condition $\delta k^\mu(-\infty)=0$,
we integrate the linearized geodesic equation (\ref{eq:geodesic_pert})
to obtain the 4 momentum
of the light ray in the {\it out} state as:
\begin{eqnarray}
\delta k^\mu(+\infty) &=&
\left\{
  \begin{array}{ll}
-2 \int_{-\infty}^{+\infty}\!d\lambda\,\partial_{||}\Psi/c^2 & (\mu=0)
   \\
 0   & (\mu=1)
  \\
-2\int_{-\infty}^{+\infty}\!d\lambda\,\partial_\perp \Psi/c^2   & (\mu=2) 
  \end{array}
\right. .
\end{eqnarray}
In the small angle scattering limit ($|\Psi/c^2|\ll 1$), the bending angle
$\hat\alpha$ of the light ray is obtained as
\begin{equation}
\hat\alpha \approx \frac{k^{\perp}(+\infty)}{k^{||}(-\infty)}
\approx  
 -\frac{2}{c^2}\int_{-\infty}^{+\infty}\!d\lambda\,  
  \partial_\perp \Psi(\lambda). 
\end{equation}
For $\mu=1$ (the $x^{||}$-component), $\delta k^{||}(\lambda)={\rm
const.}=0$, so that $k^{||}(\lambda)=1$ and 
$x^{||}=\lambda +O(|h|)$.
This suggests a simple approximation
 for the bending angle consistent
with the weak field limit: Using $d\lambda=d x^{||}+ O(|h|)$ and
integrating along the unperturbed path, we have
\begin{equation}
\label{eq:bending}
\hat{\balpha}
\approx  
 -\frac{2}{c^2}\int_{-\infty}^{+\infty}\!dx^{||}\,  
  \bnabla_\perp \Psi(x^{||},\bx_{\perp}). 
\end{equation}
This is known as the Born approximation.
This yields an explicit expression for the bending angle 
as $\hat\alpha\approx 4GM/(bc^2)\approx
1.75'' (M/M_\odot)(b/R_\odot)^{-1}$. 
General relativity gives a deflection angle
twice as large as that Newtonian physics would provide.
Eddington confirmed the prediction of general relativity 
from the measurement of the deflection of the starlight during a solar
eclipse. This solar deflection angle  is verified within $\sim 0.1\%$\cite{1995PhRvL..75.1439L}.

Finally, we find from the null condition (equation [\ref{eq:null}])
that $\delta k^0(
\lambda)=-2\Psi(\lambda) + O(h^2)$, 
or $dt/d\lambda=1-2\Psi(\lambda) + O(h^2)>1$.
The gravitational time delay $\Delta t_{\rm grav}$,
with respect to the unperturbed light
propagation, is thus given by
\begin{equation}
c\Delta t_{\rm grav} = -2\int_{-\infty}^{+\infty}\!d\lambda
 \Psi(\lambda)/c^2. 
\end{equation}


Let us consider a situation as illustrated in Figure \ref{fig:lens}:
A light ray propagates from a source ${\sl S}$ at the position $\boldeta$
in the source plane to an observer ${\sl O}$, passing the position $\bxi$
in the lens plane where it is deflected by a bending angle $\hat{\balpha}$.
Here the source and lens planes are defined as planes perpendicular to
the optical axis at the distance of the source and the lens,
respectively. The exact definition of the optical axis does not matter
since the angular scales involved in a typical lensing situation are very
small. The angle between the optical axis and the {\it true} source
position is $\bbeta$, and the angle between the optical axis and the image
$I$ is $\btheta$. The angular diameter distances from the observer to
the deflector, from the observer to the source, and from the deflector
to the source, are denoted by $D_d, D_s$, and $D_{ds}$, respectively. 
From Figure \ref{fig:lens}, we find the following geometrical relation:
$\boldeta = (D_s/D_d)\bxi - D_{ds}\hat{\balpha}(\bxi)$. Equivalently,
this is transformed to the relation between the angular source and image
positions, $\bbeta=\boldeta/D_s$ and $\btheta=\bxi/D_d$:
\begin{equation} 
\label{eq:lenseq}
\bbeta = \btheta +\frac{D_{ds}}{D_s} \hat{\balpha}(D_d\btheta)\equiv
 \btheta + \balpha(\btheta),
\end{equation}
where we have defined the reduced deflection angle $\balpha(\btheta)$ in
the last step. Equation (\ref{eq:lenseq}) is called the {\it lens
equation}, or ray-tracing equation. The lens equation is in general
non-linear (with respect to $\btheta$), so that it may have several
solutions $\btheta$ for a given source position $\bbeta$, corresponding
to the multiple imaging of a single source on the sky.


\subsection{Cosmological Lens Equation}

Here we turn to the cosmological lens equation that
describes the light propagation in an expanding, locally-inhomogeneous
universe.
There are various approaches to derive the cosmological
lens equation\cite{1985A&A...143..413S,1993PThPh..90..753S,1994CQGra..11.2345S,1995PThPh..93..647F}. 
Here we follow the approach by 
Futamase~\cite{1995PThPh..93..647F}
based on the linearized null geodesic equation as introduced in \S
\ref{sec:gl_sw}. 

Consider the following perturbation of the
Friedman-Lemaitre-Robertson-Walker (FLRW) metric:
\begin{eqnarray}
\label{eq:metric_flrw}
ds^2&=&a^2(\eta) d\tilde s^2 \equiv a^2(\eta) \tilde g_{\mu\nu} dx^\mu dx^\nu,\\
&=&a^2\left[
-(1+2\Psi)d\eta^2 + 
(1-2\Psi)
\left\{
 d\chi^2+r^2(\chi)(d\theta^2+\sin^2\theta d\phi^2)
 \right\}
\right],
\end{eqnarray}
where $(x^\mu)=(\eta,\chi,\theta,\phi)$,
$\eta=c\int^t\!dt'/a(t')$ is the conformal time, and
 $r(\chi)$ is the comoving angular diameter distance defined by
\begin{eqnarray}
\label{eq:r}
r(\chi) &=&
\left\{
  \begin{array}{ll}
K^{-1/2} \sin\left(\chi/\sqrt{K}\right) & K>0
   \\
 \chi   & K=0
  \\
(-K)^{-1/2} \sinh\left(\chi/\sqrt{-K}\right) & K<0
  \end{array}
\right. .
\end{eqnarray}
The curvature parameter $K$ is expressed with the present-day
total density parameter of the Universe
$\Omega_0$
as $K=(\Omega_0-1)H_0^2/c^2$.
Since the structure of a light cone is invariant under the conformal
transformation, we work in the conformally-related spacetime with a null
geodesic given by $d\tilde s^2$. The metric $\tilde g_{\mu\nu}$ can be
rewritten in the form of $\tilde g_{\mu\nu}=\tilde
g^{(b)}_{\mu\nu}+h_{\mu\nu}$, as a 
sum of the background FLRW metric and a small perturbation ($|h|\ll 1$).
The cosmological Poisson equation relates the Newtonian gravitational
potential $\Psi$ to the density perturbation field, $\delta\rho_m$, as:
\begin{equation} 
\label{eq:poisson}
\nabla^2 \Psi(\bchi)
=4\pi G a^2 \delta \rho_m =\frac{3H_0^2\Omega_{m}}{2}\frac{\delta_m}{a}
\end{equation}
where $\delta_m=\delta\rho_m/\bar{\rho}$ is the overdensity
with respect to 
the mean background density of the universe, 
$\Omega_{m}$ is the present-day matter density parameter,
and $\bar{\rho}$ is the mean cosmic matter density,
$\bar{\rho}=\Omega_{m} a^{-3} \rho_{\rm crit,0} = a^{-3}
(3H_0^2\Omega_{m})/(8\pi G)$.
The physical implication of equation (\ref{eq:poisson}) is that the
amplitude of $\Psi$ is related to the amplitude of $\delta$ as
$|\Psi/c^2| \sim (3\Omega_{m}/2)(l/L_H)^2 (\delta_m/a)$ where $l$ and
$L_H=c/H_0$ 
denote the
characteristic comoving scale of the density fluctuation and
the Hubble radius, respectively. 
Therefore, assuming the standard power spectrum of
the density perturbation, we can safely conclude that the degree of metric
perturbation is always much smaller than unity, i.e., $|\Psi/c^2|\ll 1$,
even for highly non-linear perturbations with $|\delta_m|\gg 1$ on
small scales of $l\ll L_H (\simeq 3\,{\rm Gpc}\, h^{-1})$.

Following the prescription given in \S \ref{sec:gl_sw}, we solve the
null geodesic equation (equations [\ref{eq:4mom}], [\ref{eq:null}],
[\ref{eq:geodesic}]) on the perturbed FLRW metric
(\ref{eq:metric_flrw}). 
Choosing the origin of the coordinate system at the observer's point
and 
backward ray-tracing from the observer $\lambda=0$ to the source
$\lambda=\lambda_s$  along the photon path,
we
obtain the 4-momentum of the light ray in the background FLRW spacetime
as $k^{(b)\mu}=(-1,1,0,0)$, and the corresponding
orbit as
$x^{(b)\mu}(\lambda)=(-\lambda,\lambda,\theta_I,\phi_I)$
with $\theta_I\ll 1$,
where $(\theta_I,\phi_I)=(\theta(0),\phi(0))$
denotes the angular direction of the image on the celestial sphere.
The comoving angular diameter distance $r(\chi)$ defined by equation
(\ref{eq:r}) can be parametrized by the affine parameter along the
photon path as $r(\chi(\lambda))=r(\lambda)$.
The null geodesic equation for the perturbative part is given by 
equations (\ref{eq:4mom_pert}), (\ref{eq:null_pert}),
(\ref{eq:geodesic_pert}), and can be formally solved as
\begin{equation}
\delta k^\mu (\lambda) =
 -\frac{2}{r^2(\lambda)}\int_0^{\lambda_s}\! d\lambda'
 \partial^\mu \Psi(\lambda')/c^2 \ \ \ (\mu=\theta,\phi),
\end{equation}
where $\partial^\mu \Psi=(\Psi_{,\theta}, \sin^{-2}\theta
\Psi_{,\phi})$. Inserting this in equation (\ref{eq:4mom_pert}) and
integrating the angular part yields the following equation:
\begin{eqnarray}
\label{eq:cosmo_lenseq}
\theta^\mu_S \equiv \theta^\mu(\lambda_s) = \theta^\mu_I - 
\frac{2}{c^2}\int_0^{\lambda_s}\!d\lambda\, 
\frac{r(\lambda_s-\lambda)}
     {r(\lambda_s)r(\lambda)}\partial^\mu \Psi(\lambda),
\end{eqnarray} 
where quantities with the subscript $S$ denote those defined for the
source.  Now we consider a small patch of the sky over which the
curvature of the sky is negligible (flat-sky approximation). Then, one
can define locally flat-sky Cartesian coordinates 
$\btheta=(\theta,\theta\phi)$ around the line-of-sight of interest
($\theta\ll 1$). 
Defining $\beta\equiv \btheta_S$ and $\btheta\equiv \btheta_I$,
equation
(\ref{eq:cosmo_lenseq}) is written in this coordinate system as
\begin{eqnarray}
\label{eq:cosmo_lenseq2}
\bbeta-\btheta
&=&\int_{\rm Observer}^{\rm Source}d\balpha
=\balpha(\chi_s),\\
\balpha(\chi_s)&=&-\frac{2}{c^2}\int_{0}^{\lambda_s}\!d\lambda\,
\frac{r(\lambda_s-\lambda)}{r(\lambda_s)} 
\bnabla_{\perp}\Psi(x(\lambda)); \ \ x(\lambda)=x^{(b)}(\lambda)+\delta x(\lambda),
\end{eqnarray}
where $\bnabla_\perp$ is the (comoving) transverse derivative,
$\bnabla_\perp\equiv
r^{-1}(\lambda)(\partial_\theta,\theta^{-1}\partial_\phi)$, and
the integral is performed along the perturbed trajectory
$x(\lambda)=x^{(b)}(\lambda)+\delta x(\lambda)$
with $\lambda=\chi + O(|\Psi/c^2|)$.
This is a general expression of the cosmological lens equation: as long
as the approximations adopted are valid 
(see Refs.~\cite{1995PThPh..93..647F} and 
\cite{takada_phd} for more details), 
equation (\ref{eq:cosmo_lenseq2}) can be applied
to various lensing phenomena, including the multiple deflections of light
rays, strong and weak lensing by clusters of galaxies, and the weak
lensing by the cosmic large-scale structure continuously distributed
between the source and the observer. Note that the cosmological lens
equation (\ref{eq:cosmo_lenseq2}) is derived with the standard angular
diameter distance in FLRW spacetime {\it without} employing the thin
lens approximation.

\subsection{Cluster Gravitational Lensing}

Now, let us turn to the case of gravitational lensing by clusters of
galaxies. Clusters are the largest gravitationally-bound systems
observed in the Universe, with typical masses of $M\sim
10^{14}-10^{15}M_\odot$. 
In the context of the standard structure formation scenario,
clusters, as the most massive collapsed structures, correspond to rare
peaks in the primordial density field. 
Clusters produce deep gravitational
potential wells, and act as efficient gravitational lenses.
The large-scale structure in the Universe also affects the
propagation of light rays from distant sources, producing small but
continuous transverse excursions of light rays along the light path.
In the study of cluster gravitational lensing, it is often assumed that
the total deflection angle $\balpha$ is dominated by the gravitational
potential of the cluster itself, and
that contributions from the cosmic large-scale structure and multiple
deflections by other non-linear objects are negligible. 
Assuming that
the light propagation is approximated by a single scattering event
by the cluster (single lens approximation) and that  
a light deflection occurs within a sufficiently small region
($\chi_d-\Delta\chi/2, \chi_d+\Delta\chi/2$) 
compared to
the relevant angular diameter distances 
(thin-lens approximation),
we have the deflection angle by a single cluster as
\begin{equation}
\balpha(\btheta) \approx -\frac{2}{c^2} \frac{D_{ds}}{D_s}
\int_{\chi_d-\Delta\chi/2}^{\chi_d+\Delta\chi/2}\!d\chi\,
\bnabla_{\perp}\Psi(\chi,r(\chi_d)\btheta),
\end{equation}
where $D_s=a(\chi_s)r(\chi_s)$ and $D_{ds}=a(\chi_s)r(\chi_s-\chi_d)$ 
are the angular diameter distances from the
observer to the source, and from the deflector to the source,
respectively, and $r(\chi_d)\btheta$ is the comoving
transverse vector on the lens plane.
In a cosmological situation, 
the angular diameter distances $D_{ij}$
between the planes $i$ and $j$ ($z_i < z_j$)
are of the order of the Hubble radius, 
$L_H\equiv c/H_0\approx 3\, {\rm Gpc}\, h^{-1}$, while physical extents
of clusters are 
about $2R=2-5\, {\rm Mpc}\, h^{-1}$. Therefore, one can
safely adopt the thin-lens approximation in cluster gravitational
lensing. 

\medskip

In actual observations, large scale structure along the line-of-sight
also contributes to the lensing signal, and consequently affects the
measurements of cluster mass properties. The level of uncertainties on
cluster lensing measurements due
to large scale structure can be assessed by numerical
simulations (see Ref.~\cite{2003MNRAS.339.1155H}).
For a massive cold dark matter (CDM)
halo with $M_{200}=10^{15}M_\odot h^{-1}$, uncertainties in the mass
measurement could reach $16-18\%$ at intermediate redshifts of
$z_d=0.2-0.3$\cite{2003MNRAS.339.1155H}.
See also Appendix \ref{appendix:multi} for the multiple
lens equation 
based on a discretized version of the cosmological lens equation.

\medskip



Then, we introduce  the {\it effective lensing potential}
$\psi(\btheta)$ defined as
\begin{equation}
\psi(\btheta) \approx
\frac{2}{c^2}
\frac{ D_{ds}}{D_{d}D_{s}}
\int_{\chi_d-\Delta\chi/2}^{\chi_d+\Delta\chi/2}
\Psi(\chi,r(\chi_d)\btheta) \, a d\chi,
\end{equation}
where $D_d$ is the angular diameter distance from the observer to the
deflector, $D_d=a(\chi_d)r(\chi_d)$.
In terms of $\psi(\btheta)$, the lens equation is expressed as
\begin{equation}
\bbeta=\btheta-\bnabla_{\theta}\psi(\btheta); \ \ \balpha(\btheta)=-\bnabla_{\theta}\psi(\btheta),
\end{equation}
where
$\bnabla_\theta=r\bnabla_{\perp}=(\partial_\theta,\theta^{-1}\partial_\phi)$. 
Table 1 gives examples of analytic axially-symmetric lens models
based on spherically-symmetric mass distributions, such as
the point mass, the singular isothermal sphere, and the NFW models.
For a more complete {\it catalog} of mass models for gravitational
lensing, see Keeton~\cite{Keeton2001}.

\begin{table}
\caption[Examples of circularly symmetric lens models]
{Examples of analytic lens models based on the spherical symmetric
 mass distribution. The 3D density profile $\rho(r)$,
 the lensing convergence $\kappa(\theta)$, and the mean convergence
 $\bar{\kappa}(\theta)$ inside the angular radius $\theta$ are given.
Here `T' and `R' indicates that the lens has a tangential and a radial 
critical curve, respectively.
} 
{\small
\begin{center}
\setlength{\doublerulesep}{2pt}
\renewcommand{\arraystretch}{2.0}
\begin{tabular}{|l|c|c|c|c|}
\hline\hline
Model & $\rho(r)$ & $\kappa(\theta)$ 
 & $\bar{\kappa}(\theta)$& Remarks \\
 \hline\hline
 Point mass & $\displaystyle M\delta_D^3(\vec{r})$ 
&$\displaystyle \frac{4\pi GM}{c^2}\frac{D_{\rm d}D_{\rm ds}}{D_{\rm s}}\delta_D^2(\vec{\theta})$ 
&$\displaystyle \frac{4 GM}{c^2}\frac{D_{\rm d}D_{\rm ds}}{D_{\rm s}}\frac{1}{\theta^2}$  
& T \\
\hline
SIS$^{(1)}$ & $\displaystyle\frac{\sigma_{v}^2}{2\pi G r^2}$ 
& $\displaystyle 2\pi G\left(\frac{\sigma_v}{c}\right)^2 
\frac{D_{\rm ds}}{D_{\rm s}}\frac{1}{\theta}$ 
& $\displaystyle 4\pi G\left(\frac{\sigma_v}{c}\right)^2 
\frac{D_{\rm ds}}{D_{\rm s}}\frac{1}{\theta}$  & T  \\
\hline
ISC$^{(2)}$ & $\displaystyle\frac{\rho(0)}{1+(r/r_{\rm c})^2}$ 
& $\displaystyle \frac{\kappa(0)}{\sqrt{1+(\theta/\theta_{\rm c})^2}}$ 
& $\displaystyle 2\kappa(0) \frac{\sqrt{1+(\theta/\theta_{\rm c})^2}-1}
{(\theta/\theta_{\rm c})^2}$& T, R 
\\
\hline
NFW$^{(3)}$ &
$\displaystyle \frac{\rho_{\rm crit}\delta_{\rm c}}
                    {(r/r_{\rm s})(1+r/r_{\rm s})^2}$ 
& $\displaystyle \kappa_{\rm s}f(\theta/\theta_{\rm s})$ 
& $\displaystyle 2\kappa_{\rm s}g(\theta/\theta_{\rm
 s})/(\theta/\theta_{\rm s})^2$ & T, R \\
\hline\hline
\end{tabular}\
\end{center}
\noindent
{\it Notes}.\\
(1) Singular isothermal sphere: $\sigma_{v}$ represents
the isothermal 1D velocity dispersion.\\
(2) Isothermal sphere with a finite core radius: 
$r_{\rm c}$ represents the core radius, and
 $\theta_{\rm c}$ is its angular radius, 
$\theta_{\rm c}:=r_{\rm c}/D_{\rm d}$; 
$\kappa(0):=\Sigma_m(0)/\Sigma_{\rm crit}$ with $\Sigma_m(0)=\pi\rho(0)r_{\rm c}$.   
 The ISC produces two critical curves if, and only if, 
$\kappa(0)>1$.\\
(3) Navarro, Frenk, and White (NFW\cite{1997ApJ...490..493N}) universal
profile of cold dark matter (CDM) halos:
$\delta_{\rm c}$, $\rho_{\rm crit}$, and $r_{\rm s}$
 represent the characteristic overdensity of the CDM halo, 
the critical density of the Universe, and the scale radius,
 respectively. $\kappa_{\rm s}$ and $\theta_{\rm s}$ are then defined by
$\kappa_{\rm s}:=2\delta_{\rm c}\rho_{\rm crit}r_{\rm s}\Sigma_{\rm crit}^{-1}$
and $\theta_{\rm s}:=r_{\rm s}/D_{\rm d}$, respectively.   
The functions $f(x)$ and $g(x)$ are defined as follows\cite{1996A&A...313..697B}\cite{2000ApJ...534...34W}:
\begin{eqnarray}
\label{eq:f(x)NFW}
f(x) &=&
\left\{
  \begin{array}{ll}
  \frac{1}{1-x^2}
   \left(
    -1+ \frac{2}{\sqrt{1-x^2}} {\rm arctanh}\sqrt{\frac{1-x}{1+x}}
   \right)  & \ \ \ (x<1) \\
\frac{1}{3} & \ \ \ (x=1); \\
  \frac{1}{x^2-1}
  \left(
   +1- \frac{2}{\sqrt{x^2-1}} \arctan\sqrt{\frac{x-1}{x+1}}  
  \right)  & \ \ \ (x>1)
  \end{array}
\right. \\
%
\label{eq:g(x)NFW}
g(x) &=& 
\ln\left(\frac{x}{2}\right)
+
\left\{
  \begin{array}{ll}
  \frac{2}{\sqrt{1-x^2}} {\rm arctanh}\sqrt{\frac{1-x}{1+x}}
   & \ \ \ (x<1) \\
 1 & \ \ \ (x=1).\\
  \frac{2}{\sqrt{x^2-1}} \arctan \sqrt{\frac{x-1}{x+1}} 
   & \ \ \ (x>1)
  \end{array}
\right. 
\end{eqnarray}  
}
\label{tab:spher_model}
\end{table}


\subsection{Lensing Jacobian Matrix}

\begin{figure}[htb]
\begin{center}
\includegraphics[width=100mm]{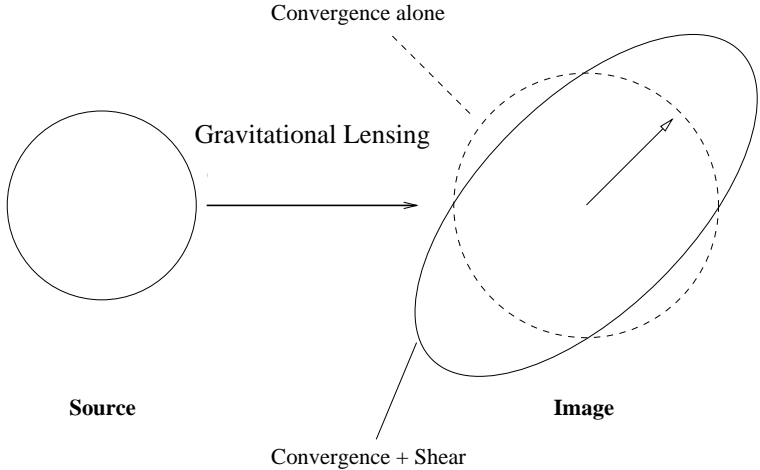}
\end{center}
\caption{
\label{fig:deformation} 
Illustration of the effects of the convergence $\kappa$ and 
the shear $\gamma$ on 
 the shape and size of a hypothetical circular source. 
The convergence acting alone causes an
 isotropic magnification of the image (dashed circle), 
while the shear deforms it to an ellipse. 
}
\end{figure}

The local properties of the lens mapping are described by the Jacobian
matrix ${\cal A}:$
\begin{equation}
{\cal A}(\btheta)=\left(\frac{\partial \bbeta}{\partial \btheta}\right)
=
\left( 
	\begin{array}{cc} 
	1 - \psi_{,11}& - \psi_{,12} \\
	-\psi_{,12} & 1 - \psi_{,22}
	\end{array}\right).
\end{equation}
This symmetric $2\times 2$ Jacobian matrix can be decomposed into the
following form:
\begin{equation}
{\cal A}_{\alpha\beta}=\delta_{\alpha\beta}-\psi_{,\alpha\beta}
=(1-\kappa)\delta_{\alpha\beta} - \gamma_1\sigma_3 - \gamma_2\sigma_1 
\end{equation}
where $\kappa(\btheta)$ is the lensing convergence field defined by
\begin{equation}
\label{eq:kappa}
\kappa = \frac{1}{2}\left(
\psi_{,11}+\psi_{,22}
\right)
=\frac{1}{2}\triangle_\theta \psi(\btheta),
\end{equation}
responsible for the trace-part of the Jacobian matrix,
 $\gamma_\alpha(\btheta)$ ($\alpha=1,2$) are the components of the
complex shear field
 $\gamma(\btheta):=\gamma_1(\btheta)+i\gamma_2(\btheta)$, defined as  
\begin{eqnarray}
\gamma_1 = \frac{1}{2}\left(\psi_{,11}-\psi_{,22}\right); \ \ 
\gamma_2 = \frac{1}{2}\left(\psi_{,12}+\psi_{,21}\right)=\psi_{,12},
\end{eqnarray}
and
$\sigma_{a} (a=1,2,3)$ are the Pauli matrices that satisfy
$\sigma_{a}\sigma_{b}=i\epsilon_{abc}\sigma_{c}$.
Equation (\ref{eq:kappa}) can be regarded as the two-dimensional Poisson
equation.
Then, the Green's function (or the propagator) 
in the infinite domain (${\cal R}^2$) is
$\triangle^{-1}(\btheta,\btheta')=\ln|\btheta-\btheta'|/(2\pi)$,
so that $\psi(\btheta)=(1/\pi)\int_{{\cal R}^2}\!d^2\theta'
\ln(\btheta-\btheta') \kappa(\btheta')$.
The explicit representation of the lens Jacobian matrix is 
\begin{equation}
{\cal A}(\btheta) =
\left( 
	\begin{array}{cc} 
	1 - \kappa - \gamma_1 & - \gamma_2 \\
	- \gamma_2 & 1 - \kappa + \gamma_1
	\end{array}\right)=
(1-\kappa)\left( 
	\begin{array}{cc} 
	1  & 0 \\
	0 & 1 
	\end{array}\right)-
\left( 
	\begin{array}{cc} 
	\gamma_1  & \gamma_2 \\
	\gamma_2 &  -\gamma_1
	\end{array}\right),
\end{equation}
and it has two eigenvalues $\Lambda_{\pm}=1-\kappa\pm|\gamma|$.
In Figure \ref{fig:deformation} we illustrate the effects of the lensing
convergence $\kappa$ and the gravitational shear $\gamma$ on the shape
and size of an infenitesimal circular source.

\subsection{Lensing Convergence}

The lensing convergence $\kappa$ is essentially a distance weighted
mass overdensity projected along the line-of-sight.
We express $\kappa(\btheta)$ due to cluster gravitational lensing as
\begin{eqnarray}
\kappa(\btheta)
&=& 
\int\! dl \, (\rho_m-\bar{\rho})
\left(
 \frac{c^2}{4\pi G}\frac{D_s}{D_s D_{ds}}
\right)^{-1} 
\simeq \frac{\Sigma_m(\btheta)}{\Sigma_{\rm crit}}; \ \ dl=ad\chi,
\end{eqnarray}
where $\Sigma_m=\int\!dl\, (\rho_m-\bar{\rho})\approx \int\!dl\,\rho_m$ 
is the surface mass
density field of the lensing cluster projected on the sky, and
$\Sigma_{\rm crit}$ is the critical surface mass density of
gravitational lensing,
\begin{equation}
\Sigma_{\rm crit}= \frac{c^2}{4\pi G}\frac{D_s}{D_s D_{ds}}
\simeq 0.1 h \, {\rm g/cm^2}\, \left(\frac{\tilde d_s}
{\tilde d_d \tilde d_{ds}}\right),
\end{equation}
where $\tilde d_{ij} \equiv D_{ij}/L_H$ is the dimensionless
angular diameter distance between the planes $i$ and $j$;
$\Sigma_{\rm crit}$ depends on the lensing geometry
($z_d, z_s$) and the geometry of the Universe, 
e.g., ($\Omega_{m}, \Omega_{\Lambda}, H_0$).

\begin{figure}[htb]
\begin{center}
\includegraphics[width=100mm,angle=270]{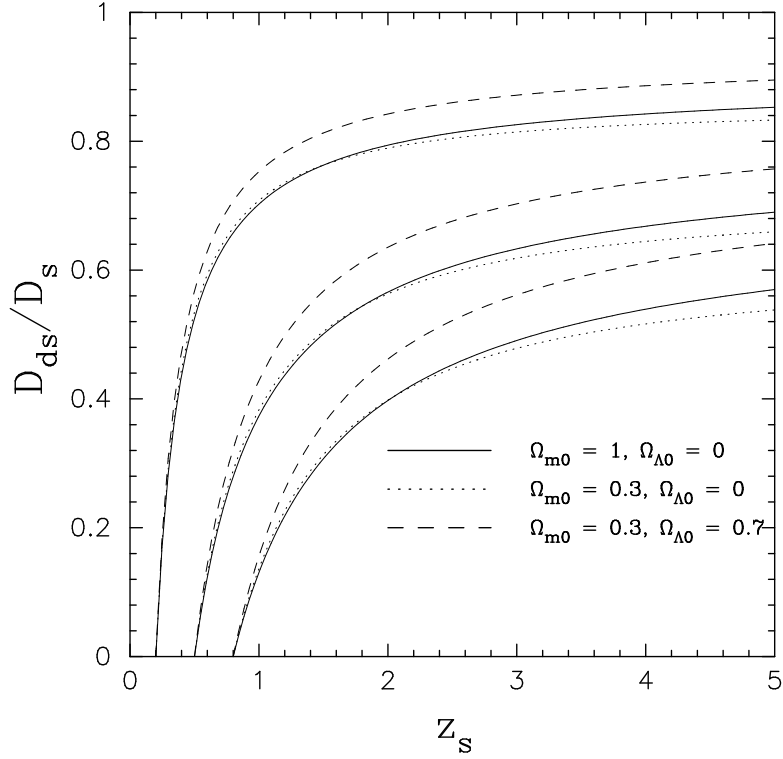}
\end{center}
\caption{
\label{fig:dratio}
Distance ratio $D_{ds}/D_{s}$ 
as a function of the
 source redshift $z_s$ for various sets of 
the lens redshift $z_d$ and the cosmological parameters
 ($\Omega_{m},\Omega_{\Lambda}$).
$D_{ds}/D_{s}$ is plotted for 
three lens redshifts $z_{d}=0.2, 0.5, 0.8$ (from left to right), and
for three sets of the cosmological parameters: 
$(\Omega_{m}, \Omega_{\Lambda})= (1,0)$, $(0.3, 0)$, and $(0.3,
 0.7)$. }
\end{figure}

\begin{figure}[htb]
\begin{center}
\includegraphics[width=80mm]{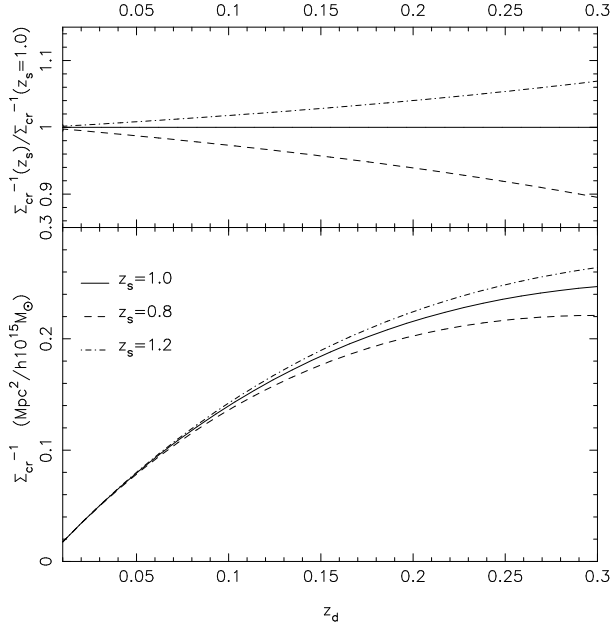}
\end{center}
\caption{
\label{fig:gleff}
Geometric scaling of the weak lensing signal.
The lower panel shows 
the inverse of the critical surface mass density of gravitational lensing,
$\Sigma_{\rm crit}^{-1} (z_d, z_s)$, 
as a function of lens redshift $z_d$
for three different source redshifts, $z_s=0.8, 1.0, 1.2$
({\it dashed}, {\it  solid}, and {\it dotted-dashed}, respectively),
demonstrating the geometric scaling of gravitational lensing signal.
The top panel shows the relative lensing strength
$\Sigma_{\rm crit}^{-1} (z_d,  z_s)/\Sigma_{\rm crit}^{-1} (z_d, z_s=1.0)$
as a function of lens redshift $z_d$
normalized with respect to the source at $z_s=1$. 
For lensing clusters at low redshifts $z_d$,
$\Sigma_{\rm crit}$ depends very weakly on the background redshift
$z_s$, 
so that the uncertainty in $z_s$ of background galaxies is less
 important in the lensing-based cluster mass determination. 
This figure is taken from Ref.~\cite{Okabe&Umetsu08}.
}
\end{figure}

For a given lens redshift $z_d$, the geometric efficiency of
gravitational lensing is determined by the distance ratio, 
$D_{ds}/D_s$, given as a function of the background redshift $z_s$ and
the cosmological parameters. Figure \ref{fig:dratio} compares
$D_{ds}/D_s$ as a function of $z_s$ for various sets of the lens
redshift and the 
cosmological model. 
In order to convert the observed lensing signal into physical mass
units, one needs to evaluate the depth of background sources
(i.e., the source redshift distribution, or its moments).
Figure \ref{fig:gleff}  shows $\Sigma_{\rm crit}^{-1}\propto
D_{ds}/D_s$ as a function of lens redshift $z_d$ for three different
source redshifts, $z_s=0.8, 1.0, 1.2$.
In typical optical imaging observations down to 
a magnitude limit of $R_{\rm c}\sim 26$ ABmag, 
the median depth of background galaxies is about $z_s\sim 1$. 
When the lens redshift is $z_d\simlt 0.2$,
$\Sigma_{\rm crit}$ depends weakly on the source redshift,
so that a precise knowledge of
the redshift distribution of background galaxies is not crucial (see for
details, e.g, Refs.~\cite{Umetsu+2010_CL0024} and
\cite{2009arXiv0903.1103O}). 
On the other hand, this distance dependence of the lensing effects can
be used to constrain the cosmological redshift-distance relation by
examining the
geometric scaling of the lensing signal as a function of the background
redshift (see Refs.~\cite{2007MNRAS.374.1377T} and
\cite{2007ApJ...663..717M}).

\subsection{Quadrupole Shape Distortion: Gravitational Shear}

The deformation of the image for an infinitesimally small background
source ($d\bbeta\to 0$)
can be described by the inverse Jacobian matrix ${\cal
A}^{-1}_{\alpha\beta}\equiv  ({\cal A}^{-1})_{\alpha\beta}$  
of the lens equation ($\alpha,\beta=1,2$).
In the weak lensing limit,
\begin{equation}
\label{eq:distortion}
{\cal A}^{-1}_{\alpha\beta} 
\approx (1 + \kappa)\delta_{\alpha\beta} + \Gamma_{\alpha\beta},
\end{equation} 
where 
\begin{equation}
\Gamma_{\alpha\beta}\equiv
\left(\partial_\alpha\partial_\beta-\delta_{\alpha\beta}
\frac{1}{2}\triangle_\theta
\right)
\psi(\btheta) =
\sigma_3\gamma_1+\sigma_1\gamma_2
\end{equation}
is the symmetric, trace-free 
$2\times 2$ shear tensor\cite{2001PhR...340..291B}\cite{2002ApJ...568...20C}.
The first term in equation (\ref{eq:distortion}) describes the isotropic
light focusing (area distortion) in the weak lensing limit, while the
second term induces an asymmetry in lens mapping; the shear is
hence responsible for the shape distortion. Note that both the
convergence and the shear contribute to the shape/area distortions in
general (non-weak) cases.

\subsection{Area Distortion: Gravitational Magnification}

The determinant of the Jacobian matrix is given as
${\rm det}{\cal A}=(1-\kappa)^2-|\gamma|^2$. 
In the weak lensing limit
where $|\kappa|, |\gamma_1|, |\gamma_2| \ll 1$, ${\rm det}{\cal A} 
\approx 1-2\kappa$.
Gravitational lensing describes the light ray deflection in the weak
field limit ($|\Psi/c^2|\ll 1$). The surface brightness of a background
source is unchanged under gravitational lensing (Liouville's theorem). 
The flux magnification
in gravitational lensing is due to the light-ray focusing that causes
the area distortion: $\delta\Omega^I=\mu\delta\Omega^S$.
The magnification is hence given by
taking the ratio between the lensed to the unlensed image solid angle
as $\mu=\delta\Omega^I/\delta\Omega^S = 1/{\rm det}A$:
\begin{equation}
\mu=\frac{1}{{\rm det}A}=\frac{1}{(1-\kappa)^2-|\gamma|^2}.
\end{equation}
In the weak lensing limit, the magnification to the first order is
\begin{equation}
\mu \approx 1+2\kappa.
\end{equation}
Thus, for an image at $\kappa(\btheta)=0.1$, the corresponding magnitude
change is $\Delta m\approx -(5/2)\log_{10}(\mu) \simeq -0.20$.
 

\section{Weak Gravitational Lensing}


In this section, we assume for simplicity that the lensing fields are
subcritical everywhere, i.e., ${\rm det}{\cal A}(\btheta)>0$.

\subsection{Weak Lensing Mass Reconstruction}
\label{sec:massrec}

\begin{figure}[htb]
\begin{center}
\includegraphics[width=80mm,angle=270]{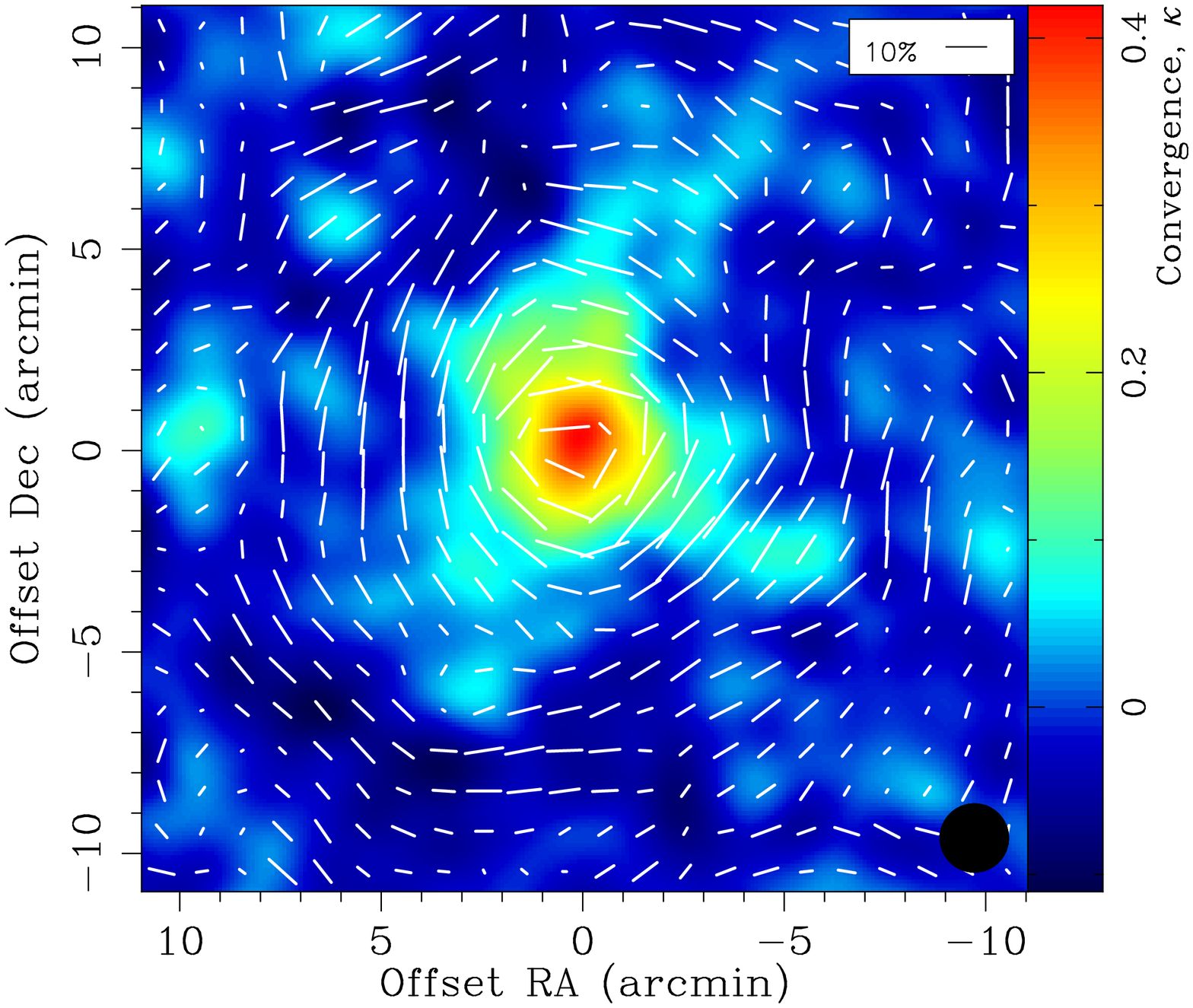}
\end{center} 
\caption{
\label{fig:mass2d}  
The projected mass distribution $\kappa(\btheta)$ of A1689 
reconstructed 
using the weak gravitational shear field $\gamma(\btheta)$
measured from a color/magnitude-selected sample of
red background galaxies registered in deep Subaru imaging observations.
Overlaid up on the image is the reconstructed spin-2 gravitational
 shear field $\gamma(\btheta)$. 
A stick with the length of $10\%$ shear is 
indicated in the top right corner.
The north is to the top, and the east is to the left.
This figure is based on the Subaru weak lensing data presented in 
Ref.~\cite{UB2008}.
}
\end{figure}

\begin{figure}[htb]
\begin{center}
\includegraphics[width=130mm,angle=0]{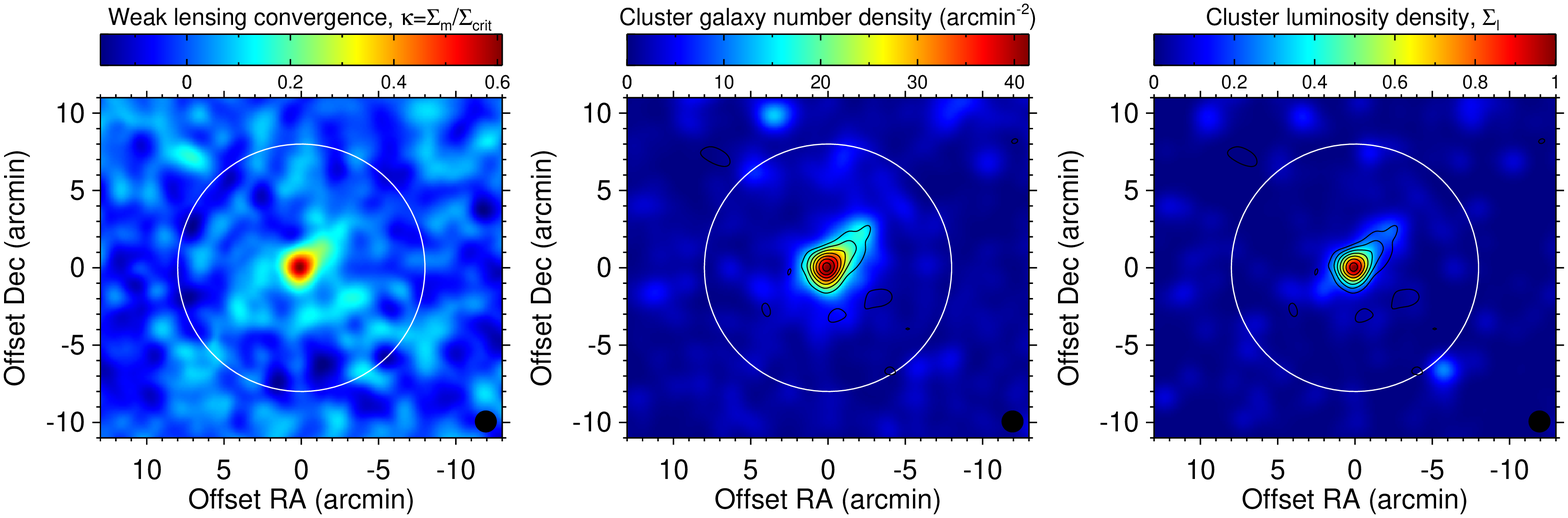}
\end{center} 
\caption{
\label{fig:3maps}
Comparison of the surface mass density field and the cluster galaxy
 distributions in Cl0024+1654. {\it Left}: Dimensionless surface mass
 density field, or the lensing convergence
 $\kappa(\btheta)=\Sigma_m(\btheta)/\Sigma_{\rm crit}$, 
reconstructed from Subaru distortion data. {\it Middle}: Observed
 surface number density distribution $\Sigma_n(\btheta)$ of $BR_{\rm
 c}z'$-selected green galaxies, representing unlensed cluster member
 galaxies. {\it Right}: Observed $R_{\rm c}$-band surface 
luminosity density distribution $\Sigma_l(\btheta)$ of the same cluster
 membership. The solid circle in 
each panel indicates the cluster virial radius of 
$r_{\rm vir}\simeq 1.8\,$Mpc\,$h^{-1}$ at the cluster 
redshift of $z=0.395$. All images are smoothed with a circular Gaussian
 of FWHM $1.4$\,arcmin. Also 
overlaid on the $\Sigma_n(\btheta)$ and $\Sigma_l(\btheta)$ maps are the
 $\kappa(\btheta)$ field shown in 
 the left panel, given in units of $2\sigma$ reconstruction error from
 the lowest contour level of $3\sigma$. The field size is $26'\times 22'$. 
North is to the top, east to the left. Figure taken from
 Ref.~\cite{Umetsu+2010_CL0024}. 
}
\end{figure}

For convenience
we define the complex gradient operator $\partial:=\partial_1+i\partial_2$
that transforms as a vector, $\partial'=\partial e^{i\varphi}$, with
$\varphi$  being the angle of rotation relative to the original basis\cite{2006MNRAS.365..414B,2007ApJ...660..995O,2008ApJ...680....1O}.
In terms of the effective lensing potential $\psi(\btheta)$, 
the lensing convergence is given as
\begin{equation}
\kappa(\btheta) = \frac{1}{2}\partial^*\partial\psi(\btheta),
\end{equation} 
where $*$ denotes the complex conjugate, and
 $\partial\partial^*=\nabla_\theta^2$ is a scalar or 
a spin-0 operator.
Similarly, the complex shear $\gamma=\gamma_1+i\gamma_2 \equiv
|\gamma|e^{2i\phi }$ is given as
\begin{equation}
\gamma(\btheta) = \frac{1}{2}\partial \partial\psi(\btheta)
\equiv \hat{\cal D}_\theta \psi(\btheta),
\end{equation} 
where $\hat{\cal D}_\theta=\partial\partial/2 =
(\partial_1^2-\partial_2^2)/2+i\partial_1\partial_2$
is a spin-2 operator, which transforms as $\hat{\cal
D}_\theta'=\hat{\cal D}_{\theta'}e^{2i\varphi}$ under a rotation of the
basis axis by $\varphi$.

Since the $\gamma$- and $\kappa$-fields are
linear combinations of the second derivatives of
$\psi(\btheta)$, $\gamma(\btheta)$ and $\kappa(\btheta)$ are related
with each other by
(see Refs.~\cite{1995ApJ...439L...1K} and \cite{2002ApJ...568...20C})
\footnote{An alternative but equivalent expression is
$\triangle_\theta
\kappa(\btheta)=\partial^{\alpha}\partial^{\beta}
\Gamma_{\alpha\beta}(\btheta)$.}
\begin{equation}
\label{eq:local}
\triangle_\theta \kappa (\btheta)
= \partial^*\partial^*\gamma
= 2\hat{\cal D}_\theta^*\gamma(\btheta)
\end{equation}
Thus, the shear-to-mass inversion formula can be formally obtained as
\begin{equation}
\label{eq:inversion}
\kappa(\btheta)
=
\triangle_{\theta\theta'}^{-1}\left[
\partial^*\partial^*\gamma(\btheta')
\right]
=2{ \hat{\cal D}^*_\theta}\triangle_{\theta\theta'}^{-1}
\left[
\gamma(\btheta')
\right].
\end{equation}
For the two-dimensional Poisson equation,
the Green's function (propagator) $\triangle^{-1}(\btheta,\btheta')$
in the infinite domain (${\cal R}^2$) is
$\triangle^{-1}(\btheta,\btheta')=\ln|\btheta-\btheta'|/(2\pi)$,
so that equation (\ref{eq:inversion}) can be solved to yield the following
non-local relation between $\kappa$ and $\gamma$ (see Ref.~\cite{1993ApJ...404..441K}):
\begin{equation}
\label{eq:gamma2kappa}
\kappa(\btheta) = 
\frac{1}{\pi}
\int_{{\cal R}^2}\!
d^2\theta'\,D^*(\btheta-\btheta')\gamma(\btheta'),
\end{equation}
where $D(\btheta)$ is the complex kernel defined as 
\begin{equation}
\label{eq:kerneld}
D(\btheta) \equiv 2\pi \hat{\cal D}_\theta \triangle^{-1}(\btheta)=
\frac{\theta_2^2-\theta_1^2-2i\theta_1\theta_2}{|\theta|^4}
=-\frac{1}{(\theta_1-i\theta_2)^2}.
\end{equation}
Similarly, the complex shear field can be expressed in terms of the
lensing convergence as
\begin{equation} 
\label{eq:kappa2gamma}
\gamma(\btheta) = 
\frac{1}{\pi}\int_{{\cal R}^2}
\!d^2\theta'\,D(\btheta-\btheta')\kappa(\btheta').
\end{equation} 
In a practical application, it is computationally fast to work in Fourier
domain\cite{2000ApJ...530..547J},
by using the fast Fourier transform (FFT). By taking the Fourier
transform of equation (\ref{eq:local}), we have a mass inversion
relation in Fourier space as
\begin{eqnarray} 
\hat{\kappa}(\bk) =\frac{k_1^2-k_2^2-2ik_1 k_2}
{k_1^2+k_2^2} \hat{\gamma}(\bk) \ \ \ (\bk \ne 0)
\end{eqnarray}
where $\hat{\kappa}(\bk)$ and $\hat{\gamma}(\bk)$
are the Fourier transform of 
the 
 $\kappa(\btheta)$ and
 $\gamma(\btheta)=\gamma_1(\btheta)+i\gamma_2(\btheta)$ 
 fields,
respectively. 

Figure \ref{fig:mass2d} shows the two-dimensional mass
distribution in the central $22'\times 22'$ region of  A1689
at $z_d=0.183$ reconstructed from the weak shear field\cite{UB2008},
measured from a sample of blue+red
background galaxies registered in deep $Vi'$ images
taken with the Suprime-Cam\cite{2002PASJ...54..833M} on the Subaru telescope.
Also overlaid up on the image is the gravitational
shear field $\gamma(\btheta)$ of the red background galaxies,
revealing a coherent tangential pattern around the cluster center.
In the left panel of Figure \ref{fig:3maps} 
we show the mass map for 
 CL0024+1654 ($z_d=0.395$) reconstructed from Subaru distortion data of 
$BR_{\rm c}z'$-selected blue+red background galaxies.
Also compared in Figure \ref{fig:3maps} are
member galaxy distributions $\Sigma_n(\btheta)$
and $\Sigma_l(\btheta)$
in the cluster,
Gaussian smoothed to the same
resolution of ${\rm FWHM}=1.41$\,arcmin.
Overall, mass and light are similarly distributed in the cluster.

Adding a constant mass sheet to $\kappa$ in equation
(\ref{eq:kappa2gamma}) does not change the 
shear field
$\gamma(\btheta)$ which is observable in the weak lensing limit, 
leading to the so-called {\it mass-sheet degeneracy}
based solely on shape-distortion measurements\cite{2001PhR...340..291B,1999PThPS.133...53U}.
As we shall see in \S \ref{sec:obs},
the observable quantity is not the 
gravitational shear $\gamma$ but the {\it reduced} shear,
\begin{equation} 
\label{eq:redshear}
g(\btheta)=\frac{\gamma(\btheta)}{1-\kappa(\btheta)}
\end{equation}
in the subcritical regime where ${\rm det}{\cal A}>0$
(or $1/g^*$ in the negative parity region with ${\rm det}{\cal A}<0$). 
We see that the $g$-field is invariant under the following
global transformation:
\begin{equation}
\label{eq:invtrans}
\kappa(\btheta) \to \lambda \kappa(\btheta) + 1-\lambda, \ \ \ 
\gamma(\btheta) \to \lambda \gamma(\btheta)
\end{equation}
with an arbitrary scalar constant $\lambda\ne 0$\cite{1995A&A...294..411S}.
This transformation is equivalent to scaling 
the Jacobian matrix ${\cal A}(\btheta)$ with $\lambda$, 
$\cal {A}(\btheta) \to \lambda {\cal
A}(\btheta)$. This mass-sheet degeneracy can be unambiguously broken
by measuring the magnification effects, because the magnification $\mu$
transforms under the invariance transformation (\ref{eq:invtrans}) as
\begin{equation}
\mu(\btheta) \to \lambda^2 \mu(\btheta).
\end{equation}


\subsection{Weak Lensing Distortion Observables}
\label{sec:obs}

\begin{figure}[!htb]
 \begin{center}
  \includegraphics[width=50mm, angle=270]{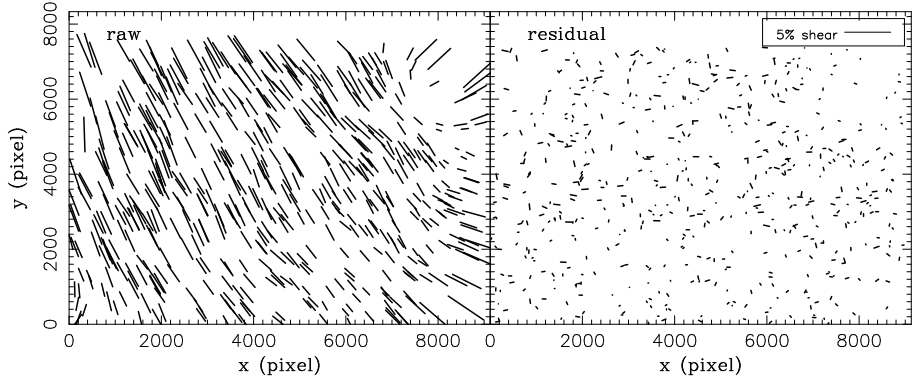}
 \end{center}
\caption{
Example of the anisotropy point-spread function (PSF) correction (Subaru
 $i'$ data of A1689).
The quadrupole PSF anisotropy field as measured from
stellar ellipticities before and after the PSF anisotropy correction.
The left panel shows the raw ellipticity field of stellar objects,
and the right panel shows the residual ellipticity field after
the PSF anisotropy correction.
The orientation of the sticks indicates the position angle of
the major axis of stellar ellipticity, whereas the length is
 proportional to the modulus of stellar ellipticity. A stick with the
 length of $5\%$ ellipticity is indicated in the top right of the right
 panel.  This figure is taken from Ref.~\cite{UB2008}.
}
\label{fig:psf1}
\end{figure}

\begin{figure}[!htb]
 \begin{center}
  \includegraphics[width=60mm, angle=270]{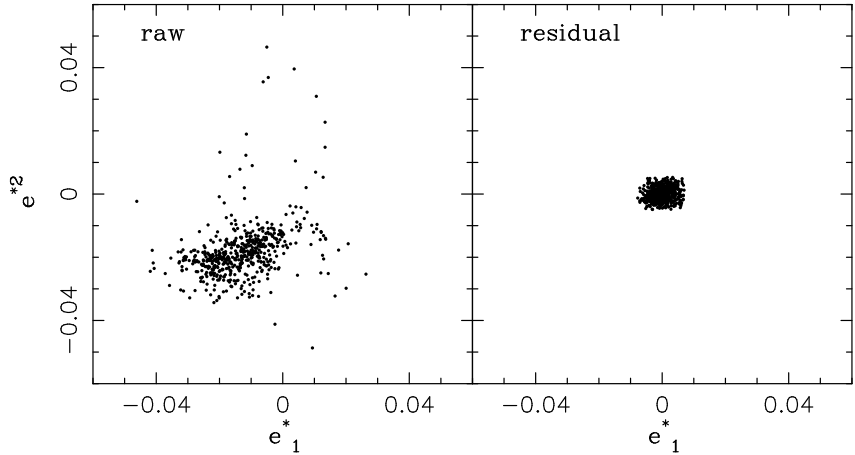}
 \end{center}
\caption{
Stellar ellipticity distributions before and after the PSF anisotropy 
correction, corresponding to Figure \ref{fig:psf1}. 
The left panel shows the raw ellipticity components 
$(e_1^*,e_2^*)$ of stellar objects, and the right panel shows
the residual ellipticity components $(\delta e_1^*, \delta e_2^*)$
after the PSF anisotropy correction. This figure is taken from
 Ref.~\cite{UB2008}.
}
\label{fig:psf2}
\end{figure}

In a moment-based approach of weak lensing shape measurements
(see Kaiser, Squires, \& Broadhurst\cite{1995ApJ...449..460K};
hereafter KSB)
we use quadrupole moments $Q_{\alpha\beta}$ ($\alpha,\beta=1,2$)
of the surface brightness distribution $I(\btheta)$ of background images
for quantifying the shape of the images:
\begin{equation}
\label{eq:Qij}
Q_{\alpha\beta} \equiv \frac{\int\! d^2\theta\, 
q_I[I(\btheta)]\Delta\theta_\alpha \Delta\theta_\beta}
{\int d^2\theta\,q_I[I(\btheta)]},
\end{equation}
where $q_I[I(\btheta)]$ denotes the weight function used in 
noisy shape measurements and $\Delta\theta_\alpha = \theta_\alpha
-\bar{\theta}_\alpha$
is the offset vector from the image centroid.\footnote{A practical
implementation of the KSB method is achieved by the IMCAT package
developed by Nick Kaiser. Note that IMCAT measures the shape moments
with respect to the peak position rather than the centroid.}
The complex ellipticity $e$ is then defined as
\begin{equation}
\label{eq:cellip}
e \equiv \frac{Q_{11} - Q_{22} + 2iQ_{12}}{Q_{11} + Q_{22}},
\end{equation}
The spin-2 ellipticity $e$ transforms under the lens mapping as
\begin{equation}
\label{eq:chis2chi}
e^{(s)}=\frac{e-2g+g^2 e^*}
                {1+|g|^2-2\Real{g e^*}},
\end{equation}
where quantities with subscript ``$(s)$'' represent 
those of (unlensed) intrinsic background sources,
and
$g = \gamma/(1 - \kappa)$ is the spin-2 reduced shear.
Since $e$ is a non-zero spin quantity
with a direction dependence, the expectation value of the intrinsic
source ellipticity $e^{(s)}$ is assumed to vanish: $\langle
e^{(s)}\rangle=0$.  
Schneider \& Seitz\cite{1995A&A...294..411S}
showed that $0=\langle \chi^{(s)}\rangle$ is equivalent to 
\begin{equation}
0=\sum_j w_j \frac{e_j - \delta_g}{1-\Re(\delta_g e^*_j)},
\end{equation}
where $\delta_g$ is the spin-2 complex distortion
$\delta_g = 2g/(1+|g|^2)$\cite{1995A&A...294..411S},
$e_j$ is the complex ellipticity for the $j$th object,
and $w_j$ is a
statistical weight for the $j$th object.
Thus, for a
intrinsically circular source with $e^{(s)}=0$, we have
\begin{equation}
\label{eq:e2delta}
e = \frac{2g}{1+|g|^2}.
\end{equation}
In the weak lensing limit ($|\kappa|, |\gamma|\ll 1$),
equation (\ref{eq:chis2chi}) reduces to  
$e^{(s)} \approx e-2g \approx e-2\gamma$.
Assuming the random orientation of the background sources,
we average observed ellipticities over a sufficient number of
images to obtain
\begin{equation}
\label{eq:chis2chiap}
\langle e\rangle\approx 2g \approx 2\gamma.
\end{equation} 
For $g^{(\rm true)}=0.1$, the weak lensing approximation (\ref{eq:chis2chiap}) 
gives $g^{(\rm est)}=0.099$, or a negative bias of $-1\%$.
For $g^{(\rm true)}=0.4$ in the non-linear regime, 
equation (\ref{eq:chis2chiap}) 
gives $g^{(\rm est)}=0.345$, corresponding to a negative bias of $-14\%$.

For a practical application of weak lensing shape measurements,
we must take into account various observational effects
such as noise in the shape measurement due to readout
and/or sky background
and the dilution of the lensing signal due to the isotropic/anisotropic
point-spread function (PSF) effects.
Thus, one cannot simply use equation
(\ref{eq:chis2chiap})
to measure the gravitational shear field.
KSB took into account
explicitly the Gaussian weight function in
calculations of noisy shape moments and the effect of 
quadrupole PSF anisotropy, as well as isotropic PSF smearing,
and derived in the limit of linear anisotropies
 relevant
transformation equations between 
unlensed (intrinsic) and lensed (observed) ellipticities.
In the limit of linear anisotropies, the transformation equation between
the intrinsic  and the observed complex ellipticities
is
formally expressed as 
 \begin{eqnarray}
e_\alpha  &=&  e_\alpha^{(s)} + 
\left(C^{g}\right)_{\alpha\beta} g_\beta 
+ 
\left(C^q\right)_{\alpha\beta} q_\beta,
\end{eqnarray}
where
$q_\alpha$ is the spin-2 PSF anisotropy kernel,
and
$C$s 
are linear response coefficients for the spin-2 anisotropy fields
($g_\alpha$ and $q_\alpha$),
which can be calculated from observable
weighted shape moments of galaxies and stellar objects\cite{2001PhR...340..291B,1995ApJ...449..460K,2001A&A...366..717E}.
In practical observations, the spin-2 PSF anisotropy $q(\btheta)$ can be
measured from image ellipticities of foreground stars, for which
both of $e^{(s)}$ and $g$ vanish:
$q_\beta(\btheta)=(C^q)^{-1}_{\alpha\beta}e^*_{\beta}$.
As an example, we show in Figure \ref{fig:psf1}
the quadrupole PSF anisotropy field
as measured from stellar ellipticities before and after the anisotropic
PSF correction using the Subaru $i'$ data of the cluster A1689\cite{UB2008}.
Figure \ref{fig:psf2} shows
the distribution of stellar ellipticity
components before and after the PSF anisotropy correction, corresponding
to Figure \ref{fig:psf1}.

Assuming that the expectation value of the intrinsic source ellipticity
vanishes, we have the linear relation between 
the averaged image ellipticity and the reduced shear as
\begin{eqnarray}
g_\alpha  &\approx &  
\langle \left( C^{g}\right)^{-1}_{\alpha\beta} (e-C^q q)_\beta \rangle.
\end{eqnarray} 
In the context of the KSB formalism, the linear response
$C^g_{\alpha\beta}$ to the gravitational shear is often denoted as
$P^g_{\alpha\beta}$ (or $P^\gamma_{\alpha\beta}$). A careful calibration
of $P^g$ is crucial for accurate measurements of the weak lensing
signal. 
See Refs.~\cite{2006MNRAS.368.1323H} and \cite{2007MNRAS.376...13M}
for more detailed discussions on the shear calibration issues.

\subsection{$E$/$B$ Decomposition}
\label{sec:eb}

In general, the shear tensor field
$\Gamma_{\alpha\beta}(\btheta)=\gamma_1(\btheta)\sigma_3+\gamma_2(\btheta)
\sigma_1$   
that describes the spin-2 quadrupole shape distortions
can be expressed as a sum of two terms, corresponding to two degrees of
freedom, 
by introducing two scalar functions $\Phi_E(\btheta)$ and
$\Phi_B(\btheta)$ 
(see Ref.~\cite{2002ApJ...568...20C}) as 
\begin{eqnarray} 
\Gamma_{\alpha\beta}(\btheta)
&=&
\left( 
	\begin{array}{cc} 
	\gamma_1 & \gamma_2\\
	\gamma_2 & -\gamma_1
	\end{array}\right)
= 
\Gamma_{\alpha\beta}^{(E)}(\btheta)
 +\Gamma_{\alpha\beta}^{(B)}(\btheta), \\
\Gamma_{\alpha\beta}^{(E)}&=&
 \left(\partial_\alpha\partial_\beta-\delta_{\alpha\beta}
  \frac{1}{2}\triangle_\theta\right)\Phi_E(\btheta); \ \ 
\Gamma_{\alpha\beta}^{(B)}=\frac{1}{2}
\left(
\epsilon_{\gamma\beta}\partial_\alpha\partial_\gamma
+
\epsilon_{\gamma\alpha}\partial_\beta\partial_\gamma
\right)\Phi_B(\btheta),
\end{eqnarray}
where $\epsilon_{\alpha\beta}$ is the $2\times 2$ antisymmetric tensor,
$\epsilon_{11}=\epsilon_{22}=0$, $\epsilon_{12}=-\epsilon_{21}=1$.
The first term associated with $\Phi_E$ is  a gradient or scalar $E$
component, and the second term with $\Phi_B$ is a curl or pseudoscalar
$B$ component. The $\gamma_1$ and $\gamma_2$ are then written in
terms of $\Phi_E$ and $\Phi_B$ as
\begin{eqnarray}
\gamma_1&=&+\Gamma_{11}=-\Gamma_{22}=\frac{1}{2}\left(\Phi_{E,11}-\Phi_{E,22}\right) 
 -\Phi_{B,12}\\
\gamma_2&=&\Gamma_{12}=\Gamma_{21}=\Phi_{E,12}
+\frac{1}{2}\left(\Phi_{B,11}-\Phi_{B,22}\right).
\end{eqnarray}
As we have seen in \S \ref{sec:massrec}, the spin-2 $\gamma$ field is
coordinate dependent, which transforms as $\gamma'=\gamma e^{2i\varphi}$
under a rotation of the basis axis by $\varphi$. 
The $E$ and $B$ parts can be extracted from the shear tensor by applying
the $\nabla^4_\theta$ operator:
\begin{equation}
\label{eq:emode}
2\nabla^2_\theta\kappa_E\equiv
\nabla^4_\theta\Phi_E=2\partial^\alpha\partial^\beta\Gamma_{\alpha\beta},
\ \ 
\label{eq:bmode}
2\nabla^2_\theta\kappa_B\equiv
\nabla^4_\theta\Phi_B=2\epsilon_{\alpha\beta}
\partial^\alpha\partial^\gamma\Gamma_{\beta\gamma},
\end{equation}
where we have defined the $E$ and $B$ fields,
$\kappa_E=(1/2)\triangle_\theta\Phi_E$ and
$\kappa_B=(1/2)\triangle_\theta\Phi_B$, respectively. 
This technique is called as the $E$-$B$ mode decomposition.
The above equations tell us that
the relations between the $E$/$B$-fields and the spin-2 
field are intrinsically non-local.

Remembering that in the case of gravitational lensing
the shear tensor is given as
$\Gamma_{\alpha\beta}=(\partial_\alpha\partial_\beta-
\delta_{\alpha\beta}\triangle_\theta/2)\psi(\btheta)$, we can identify
$\Phi_E(\btheta)=\psi(\btheta)$ and $\Phi_B(\btheta)=0$. Hence, for a
lensing-induced shear field, the $E$-mode signal is related with the
convergence $\kappa$, i.e., the surface mass density of the lens,
while the $B$-mode signal is identically zero. 
We note that gravitational lensing can give rise to $B$-modes, for
example, when multiple deflections of light rays along the light path
are involved. However, these $B$ modes arise at higher orders and 
the $B$-mode contributions coming from multiple deflections are
suppressed by a large factor compared to the $E$-mode contributions.
In practical observations, intrinsic ellipticities of background
galaxies 
also contribute to the gravitational shear estimate, $\gamma^{\rm est}$.
Assuming that intrinsic ellipticities 
have random orientations in projection space, such uncorrelated
ellipticities will yield statistically identical contributions
to  the $E$- and $B$ modes.
Thus the $B$-mode signals serve as a useful null check for
systematic effects  (e.g., residual PSF anisotropies).

Now we turn to the issue of $E$/$B$-mode reconstructions from the
spin-2 shear field. Rewriting equations (\ref{eq:emode}) and
(\ref{eq:bmode}) in terms of the complex shear $\gamma$, we find
\begin{eqnarray}
\label{eq:emode_inv}
\triangle_\theta\kappa_E&=&
\Re\left(
2\hat{\cal D}_{\theta}^*\gamma\right), \\
\label{eq:bmode_inv}
\triangle_\theta\kappa_B&=&
\Im\left(
2\hat{\cal D}_{\theta}^*\gamma
\right)=
-\Re\left(
2\hat{\cal D}_{\theta}^* i\gamma\right).
\end{eqnarray}
Defining $\kappa\equiv \kappa_E+i\kappa_B$, we see that 
equations (\ref{eq:emode_inv}) and (\ref{eq:bmode_inv}) are identical
to the mass inversion equation (\ref{eq:local}).
Therefore, the $B$-mode convergence $\kappa_B$ can be simply obtained as
the imaginary part of equation (\ref{eq:gamma2kappa}), which is expected to
vanish for a purely weak lensing signal.
Furthermore, the second equality of equation (\ref{eq:bmode_inv})
indicates that the transformation
$\gamma'(\btheta)=i\gamma(\btheta)$ 
($\gamma_1'=-\gamma_2, \gamma_2'=\gamma_1$)
is equivalent to an interchange operation of the $E$ and $B$ modes 
of the original
maps: $\kappa_E'(\btheta)=-\kappa_B(\btheta)$,
$\kappa'_B(\btheta)=\kappa_E(\btheta)$.  
Since $\gamma$ is a spin-2 field that transforms as $\gamma'=\gamma
e^{2i\varphi}$, this operation is also equivalent to a rotation of each
ellipticity by $\pi/4$ with each position vector fixed.

\subsection{Magnification Bias}
\label{subsec:magbias}

In the absence of gravitational lensing,
the cumulative number counts $n_0(>S_0)$ (per solid angle)
of background galaxies 
above the limiting flux $S_0$
can be locally approximated 
around $S=S_0$
by a power-law form:
\begin{equation}
n_0(>S_0) \equiv \int_{S_0}^{\infty}\!dS\, \frac{d^2 N}{d\Omega dS}
\propto S_0^{-\alpha}
\end{equation}
with the running power index around $S=S_0$
\begin{equation}
\alpha \equiv -\frac{d\log_{10} n_0(>S_0) }{d\log_{10} S_0} > 0.
\end{equation}
Note the value of $\alpha$ depends on the luminosity function
$dn/dL(L,z)$ of background sources (and hence the object type, such as
late/early type galaxies and  quasars) and the observing wavelength.


Gravitational lensing induces the following conflicting effects
known as  {\it magnification bias} 
(see Refs.~\cite{1995ApJ...438...49B} and \cite{1998ApJ...501..539T}): 
\begin{enumerate}
\item[1] Area distortion: $\delta\Omega^I=\mu(\btheta)\delta\Omega^S$
\item[2] Flux Magnification: $S\to \mu(\btheta)S$.
\end{enumerate}
The former effect reduces the effective observing area in the source plane,
decreasing the number of background sources per solid angle;
on the other hand, the latter effect amplifies the flux of background sources,
increasing the number of sources above the limiting flux.
The net number counts with gravitational lensing is given as
\begin{equation}
n(>S_0) \equiv \int_{S_0/\mu}^{\infty}\!dS\, \frac{d^2 N}{\mu d\Omega
 dS}
=\mu^{\alpha-1} \, n_0(>S_0).
\end{equation}
This implies that, (i) positive bias for $\alpha>1$ and (ii) negative
bias for $\alpha<1$.

In the weak lensing limit  ($|\kappa|,|\gamma_\alpha|\ll 1$),
\begin{equation}
n(>S_0) 
\approx 
(1+2\kappa)^{\alpha-1} \, n_0(>S_0)
\approx 
\left\{
1+2(\alpha-1)\kappa)
\right\}
 \, n_0(>S_0).
\end{equation}
The fractional change in the surface number density of background
sources is thus given as
\begin{equation}
\delta_N = \frac{\delta n}{n_0}
\approx 
-2(1-\alpha)\kappa.
\end{equation}
Hence the magnification-bias $\delta_N(\btheta)$ in
the weak-lensing limit is a local
measure of the surface-mass density field, $\kappa(\btheta)$.
A combination of the lens distortion and magnification can thus be used
to break the mass-sheet degeneracy inherent in the shear-based mass
determination\cite{BTU+05,UB2008}.



\section{Cluster Weak Lensing Profiles}


In this section we summarize basic,  useful
techniques for measuring cluster weak lensing profiles, which can be
compared quantitatively with various cluster models, 
and can be used to constrain cluster mass and structure parameters.

\subsection{Weak Lensing Distortion}
\label{sec:gtx}

The spin-2 shape distortion of an object 
due to gravitational lensing
is described by
the complex reduced shear, $g=g_1+i g_2=\gamma/(1-\kappa)$, which however 
is coordinate dependent.
For a given reference point on the sky, one can instead 
form coordinate-independent
quantities, 
the tangential distortion $g_+$ and the $45^\circ$ rotated component,
from linear combinations of the distortion coefficients
$g_1$ and $g_2$ as\cite{1995ApJ...446L..55T}
\begin{equation}
g_+ = -(g_1 \cos 2\varphi + g_2\sin 2\varphi), \ \ 
g_{\times} = -(g_2 \cos 2\varphi - g_1\sin 2\varphi),
\end{equation}
where $\varphi$ is the position angle of an object with respect to
the reference position, and the uncertainty in the $g_+$ and
$g_{\times}$ 
measurement 
is $\sigma_+ = \sigma_{\times } = \sigma_{g}/\sqrt{2}\equiv \sigma$ 
in terms of the rms error $\sigma_{g}$ for the complex distortion
measurement.
The $+$-component, $g_+$, is a measure of tangential coherence of the
shape distortions of background images due 
weak gravitational lensing  
(see Figure \ref{fig:mass2d} and discussions in \S \ref{sec:eb}) . 
On the other hand, the
$\times$-component, $g_\times$, corresponds to divergence-free,
curl-type distortion patterns of background images.
In practice, the reference point is taken to be the cluster center,
which can be determined from symmetry of the strong lensing pattern,
the X-ray centroid position, or the brightest cluster galaxy
position\cite{2009arXiv0903.1103O}. 
To improve the statistical significance of the distortion measurement,
we calculate the weighted average of the $g_+$'s and its weighted error 
as
\begin{eqnarray}
\label{eq:gtx}
\langle g_+(\theta_n)\rangle &=&\frac{\sum_j  u_{g,j} g_{+,j}}{\sum
 u_j}, \ \ \ 
\langle g_\times(\theta_n)\rangle =\frac{\sum_j  u_{g,j} g_{\times,j}}{\sum
 u_j}, \\
\label{eq:cov_gtx}
C_{mn}&\equiv&\langle \Delta g_{+}(\theta_m)\Delta
g_{+}(\theta_n)\rangle=
\langle \Delta g_{\times}(\theta_m)\Delta
g_{\times}(\theta_n)\rangle 
=\delta_{mn}\sigma_{+}^2(\theta_m) \\
\label{eq:err_gtx}
\sigma_+(\theta_n) &=& \sigma_\times(\theta_n)= \left(
\frac{\sum_j u_{g,j}^2\sigma_j^2}
{\left(\sum_j u_{g,j}\right)^2} 
\right)^{1/2},\\
\label{eq:rm}
\theta_n &=&
\displaystyle\sum_{j\in {\rm bin}\,n} u_{g,j}|\btheta_j|
\Big/ 
\displaystyle\sum_{j\in {\rm bin}\,n} u_{g,j},
\end{eqnarray}
where the $j$ runs over all of the objects located within the $n$th
annulus with a median radius of $\theta_n$, 
$g_{+,j}$ is the $+$-component of the reduced shear estimate for
the $j$th object, $u_{g,j}$ is a statistical weight for the $j$th
object,  and $C_{mn}$ is the bin-to-bin covariance error matrix of the
binned distortion profiles.
Here we have used $\langle g_{\alpha,i}g_{\beta,j}
\rangle=(1/2)\sigma_{g,j}^2\delta_{\alpha\beta}\delta_{ij}$
($\alpha,\beta=1,2$).\footnote{We have ignored the contribution
of large scale structure along the line of
sight to the lensing signal, namely the cosmic shear signal,
acting as spatially correlated noise on
the cluster lensing measurement. See Ref.~\cite{2003MNRAS.339.1155H}.} 
Several authors adopted the statistical weight
$u_{g,i}$ of the form: $u_{g,i}=1/(\sigma_{g,j}^2+\alpha^2)$, with
$\alpha$ being the softening constant variance\cite{2003ApJ...597...98H,Okabe+Umetsu2008,Umetsu+2010_CL0024}.
In the limit of $\alpha\gg \sigma_{g}^2$, this
corresponds to the uniform weighting.  The case of $\alpha=0$ is known
as the inverse-variance weighting, yielding
$\sigma_+(\theta_n)=\sigma_\times(\theta_n)=1/\sqrt{\sum_j(1/\sigma_{j}^2)}$.  
In Refs.~\cite{2003ApJ...597...98H} and \cite{Okabe&Umetsu08}
$\alpha\sim \langle
\sigma_{g}^2\rangle\approx 0.4$ was used as the softening parameter for the
weight function.

The tangential reduced shear $g_+(\theta)$ as a function of angular radius
is a useful, direct observable in weak lensing observations, being free
from the mass-sheet degeneracy (\S \ref{sec:massrec}): It does not
require a non-local mass reconstruction (see \S \ref{sec:massrec} and
\ref{sec:eb}), and hence
one can
easily assess its error propagation as given by
equations (\ref{eq:cov_gtx}) and  (\ref{eq:err_gtx}). 
In particular, the bin-to-bin covariance error matrix $C_{mn}$ is
diagonal (assuming that statistical uncertainties are dominated by
random orientations of intrinsic ellipticities).
Furthermore, the $45^\circ$-rotated $g_\times$ component can be used
as a useful null check for systematic effects. 
To compare observed tangential-distortion profiles with cluster
mass models, one can use the identify $\langle \gamma_+ (\theta)\rangle
= \overline{\kappa}(<\theta)-\langle \kappa(\theta)\rangle$\cite{2001PhR...340..291B}\cite{1999PThPS.133...53U},
where $\langle
\gamma_+(\theta)\rangle$ is the azimuthally-averaged tangential
component of the gravitational shear at radius $\theta$, 
$\langle \kappa(\theta)\rangle=\langle\Sigma_m(\theta)
\rangle/\Sigma_{\rm crit}$ is the azimuthally-averaged $\kappa$ at
radius   
$\theta$, and
$\overline{\kappa}(<\theta)=\overline{\Sigma}_m(<\theta)/\Sigma_{\rm
crit}$ is the mean lensing 
convergence interior to radius $\theta$. If the projected cluster mass
distribution is azimuthally symmetric about its center, then one can
make an approximation $\langle g_+(\theta)\rangle\simeq 
( \overline{\kappa}(<\theta) - \langle\kappa(\theta)\rangle) 
/(1-\langle\kappa(\theta)\rangle)$ to compare with predictions from
azimuthally-symmetric mass models (see Table 1).

\begin{figure}[htb]
\begin{center}
\includegraphics[width=80mm,angle=270]{fig_gplot_a1689.ps}
\includegraphics[width=80mm,angle=270]{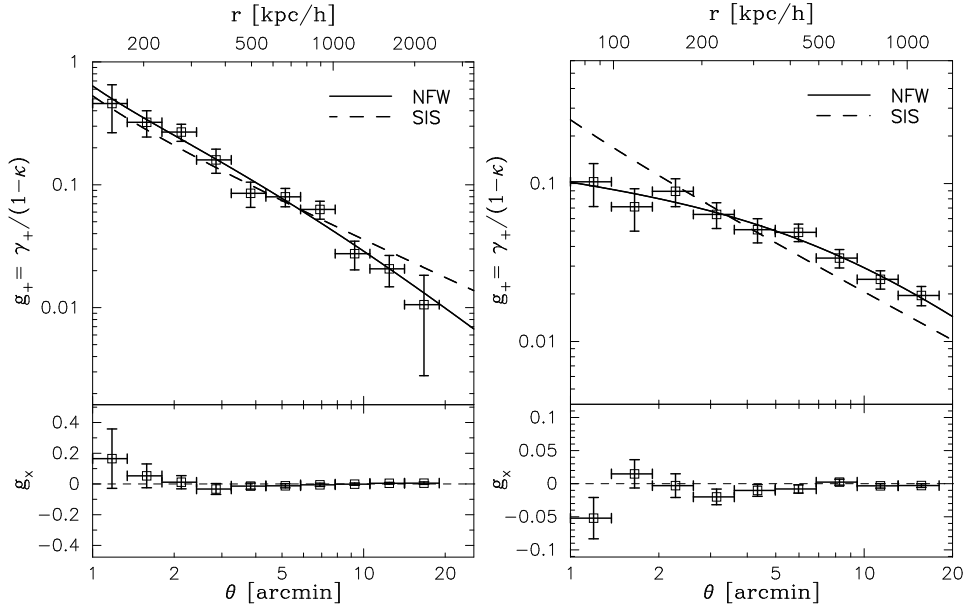}
\end{center} 
\caption{
\label{fig:gtx} 
Azimuthally-averaged radial profiles of the tangential reduced shear
 $g_+$ ({\it upper panels}) for the high-mass galaxy clusters A1689
at $z_d=0.183$ ({\it left}) and A2142 at $z_d=0.091$ ({\it right})
based on deep Subaru weak lensing data.
The solid and dashed curves show the best-fitting NFW and SIS profiles
for each cluster.
Shown below is the  $45^\circ$ rotated ($\times$) component, $g_\times$.
Figure taken from Ref.~\cite{2009ApJ...694.1643U}.
} 
\end{figure}

In Figure \ref{fig:gtx} we show the tangential and $45^ \circ$-rotated
distortion profiles for A1689 at $z_d=0.183$ ({\it left}) and A2142 at
$z_d=0.091$ 
({\it right}) derived from Subaru weak lensing
data\cite{2005ApJ...619L.143B,UB2008,2009ApJ...694.1643U}, 
along with the respective
best-fitting $g_+(\theta)$-profiles for the NFW (see Ref.~\cite{1997ApJ...490..493N}) 
and singular isothermal sphere (SIS) models (see Table 1).
In both cases, the cluster distortion profiles are better fitted by the NFW
model with a continuously steepening profile. For the nearby, high-mass
cluster A2142, the curvature in the distortion profile appears 
highly pronounced, and the SIS model can be strongly rejected by the
weak lensing data alone.

\subsection{Weak Lensing Depletion}

\begin{figure}[htb]
\begin{center}
\includegraphics[width=110mm,angle=0]{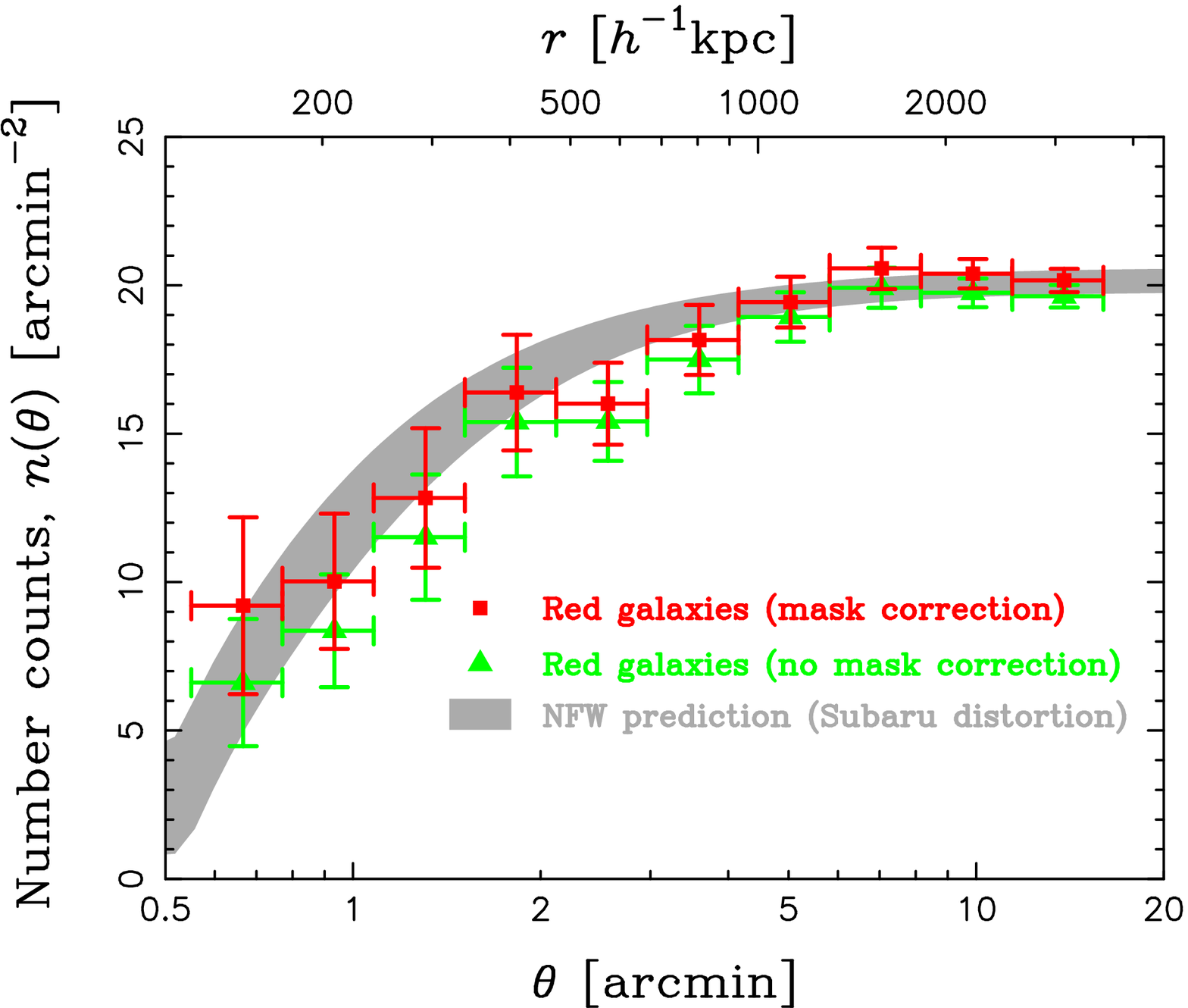}
\end{center} 
\caption{
\label{fig:depletion} 
Number-count profile of Subaru $BR_{\rm c}z'$-selected red galaxies
 ({\it squares}) 
in the background of Cl0024+1654 at $z_d=0.395$. The triangles show the
counts without the mask correction due to cluster members and bright
 foreground objects. 
A slight dip at $\theta=2'-3'$ in the depletion profile
corresponds to the contribution of the known substructure (see Figure
 \ref{fig:3maps}), which is also 
 seen in Subaru distortion data.
The gray-filled region represents the 68.3\% confidence 
bounds for the predicted count depletion curve from an NFW model
 constrained by our Subaru distortion analysis, demonstrating clear
 consistency between these two independent lensing observables. Figure
 taken from Ref.~\cite{Umetsu+2010_CL0024}.
} 
\end{figure}

Lensing magnification $\mu$
influences the observed surface density of background
sources, expanding the area of sky, and enhancing the observed flux of
background sources (\S \ref{subsec:magbias}).
For red background galaxies at a median redshift of 
$\overline{z_s}\sim 1$,
the intrinsic count slope 
$s=-d\log_{10}n_0(>S_0)/d\log_{10}S_0=(5/2)d\log_{10}n(<m_0)/dm_0$
at faint magnitudes $m_0$ is relatively flat ($s<1$),
so that a net count
depletion results\cite{1995ApJ...438...49B}.
Recently the count depletion of red background galaxies due to
gravitational lensing has been clearly detected in several massive clusters
(A1689, A1703, A370, RXJ1347-11,
CL0024+1654)\cite{BTU+05,BUM+08,Umetsu+2010_CL0024}.
Figure \ref{fig:depletion} displays the count-depletion profile derived
from a Subaru 
$BR_{\rm c}z'$-selected red galaxy sample in the background of CL0024+1654
at $z_s=0.395$\cite{Umetsu+2010_CL0024}.
The error bars include not only the Poisson
contribution but also the variance due to variations of the counts along
the azimuthal direction, i.e., contributions from the intrinsic
clustering of red galaxies\cite{1995ApJ...438...49B,2005PhRvL..95x1302Z} 
and departure from circular symmetry\cite{UB2008}. 
A strong depletion of the red galaxy counts is shown in the central,
high-density region of the cluster, and clearly detected out to a few
arcminutes from the cluster center. 
The gray-filled region represents the 68.3\% confidence
bounds for the predicted count depletion curve
$n(\theta)=n_0\mu^{s-1}(\theta)$ from an NFW model constrained by Subaru
distortoin data,
demonstrating clear consistency between these two independent
lensing observables.

\subsection{Weak Lensing Dilution}

It is crucial in the cluster weak lensing analysis
to make a secure selection of background galaxies in order
to minimize contamination by 
cluster/foreground galaxies and hence to make an accurate determination 
of the cluster mass,
otherwise dilution of the distortion signal results from
the inclusion of unlensed cluster galaxies, particularly at small radius
where the cluster is relatively
dense\cite{2005ApJ...619L.143B}\cite{2007ApJ...663..717M}. 

This dilution effect is 
simply to reduce the strength of
the lensing signal when averaged over a local ensemble of 
galaxies, 
in proportion to the fraction of unlensed cluster and foreground
galaxies whose orientations are randomly distributed, thus diluting the lensing
signal relative to the true background level, derived from
the uncontaminated background population\cite{2007ApJ...663..717M}. 
With a pure red background sample (B) as a reference,
one can quantify the {\it degree of dilution} for a galaxy sample (G)
containing $N_{\rm CL}$ cluster galaxies and $N_{\rm BG}$
background galaxies
in terms of the strengths of the averaged tangential shear signal
$\langle g_+(\theta)\rangle$
as\cite{2007ApJ...663..717M}
\begin{equation}
1+\delta_d(\theta) \equiv \frac{N_{\rm BG}+N_{\rm CL}}{N_{\rm BG}}
=\frac{\langle g_+^{(B)}(\theta)\rangle}
      {\langle g_+^{(G)}(\theta)\rangle}
 \frac{\langle D_{ds}/D_s\rangle^{(G)}_{z_s\ne z_d}}
      {\langle D_{ds}/D_s\rangle^{(B)}_{z_s\ne z_d}},
\end{equation}
where  $\langle D_{ds}/D_s\rangle_{z_s\ne z_d}$'s are 
averaged distance ratios 
for respective background populations\footnote{{\it
Background} samples can be generalized to include foreground field
galaxies with $z_s<z_d$ 
and $D_{ds}/D_s(z_s)=0$};  
if the two samples contain the same
background population, then 
$\delta_d = 
\langle g_+^{(B)}\rangle/\langle g_+^{(G)}\rangle-1$.
The degree of dilution thus varies depending on the radius from
the cluster center,  increasing towards the cluster center.
Medezinski et al.~\cite{2007ApJ...663..717M} 
found 
for their {\it green} galaxy sample
 ($[V-i']_{E/S0-0.3}^{+0.1}$) containing
the cluster sequence galaxies in A1689
that the fraction of cluster membership,
$N_{\rm CL}/(N_{\rm BG}+N_{\rm CL})$,
tends $\sim 100\%$ within $R \simlt 300$\,kpc$\,h^{-1}$.

\begin{figure}[!htb]
 \begin{center}
  \includegraphics[width=70mm,angle=270]{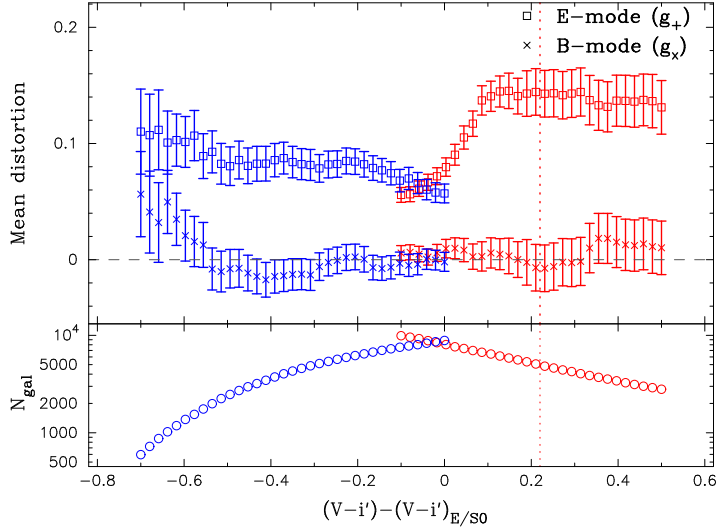}
 \end{center}
\caption{
\label{fig:dilution}
Top panel:
mean distortion strength averaged over a wide radial range of $1'
 < \theta < 18'$ of A1689,
done separately for the blue and red galaxy samples.
No area weighting is used here to enhance the effect of dilution in the
 central cluster region.
Shown are the measurements of the tangential component ($g_+$)
with open squares, and those of the $45{\rm deg}$-rotated component
 ($g_\times$).  
Bottom panel: the respective numbers of galaxies as a function of color
 limit, contained in the range
 $1' <\theta < 18'$ in the red  ({\it right}) and the blue
 ({\it left}) samples. This figure is taken from Ref.~\cite{UB2008}.
} 
\end{figure}

Broadhurst, Takada, Umetsu et al. (see Ref.~\cite{2005ApJ...619L.143B})
proposed to  
use a sample of {\it red background galaxies}
whose colors are redder due to large $k$-corrections than
the color-magnitude relation, or red sequence, of cluster member galaxies.
These red background galaxies are largely composed of early to mid-type
galaxies at moderate redshifts\cite{2007ApJ...663..717M}.
Cluster member
galaxies are not expected to extend to these colors in any significant
numbers because the intrinsically reddest class of cluster galaxies,
i.e. E/S0 galaxies, are defined by the red sequence and lie blueward of
chosen sample limit, so that even large photometric errors will not 
carry them into such a  red sample.
This can be demonstrated readily, as shown in Figure \ref{fig:dilution},  
where we plot the mean tangential shear strength $\langle g_+ \rangle$
of A1689,
averaged over a wide radial range of $1'< \theta < 18'$,
as a function of color limit
by changing the lower color limit progressively blueward.
Here we do not apply area weighting to enhance the effect of dilution in
the central region.
Figure \ref{fig:dilution} shows a sharp drop in the
lensing signal at $\Delta(V-i') \simlt 0.1$,
when the cluster red sequence starts to contribute significantly,
thereby reducing the mean lensing signal.
At $\Delta(V-i')\simgt 0.1$,
the mean lensing signal of the red background
stays fairly constant, $\langle g_+ \rangle \simeq 0.143$,
ensuring that our weak lensing measurements 
are not sensitive to this particular choice of the color limit.

Recently,
this empirical background selection method
has been generalized by Medezinski, Broadhurst, Umetsu et
al.~\cite{Medezinski+2009} 
to incorporate and combine all color and positional information in a
color-color (CC) diagram. This CC-selection method has been
successfully applied
to Subaru imaging observations of several massive
clusters\cite{Medezinski+2009,Umetsu+2010_CL0024}. 

On the other hand,
the dilution of the lensing signal caused by cluster members can be 
used to derive the proportion of galaxies statistically belonging
to the cluster, or the cluster member fraction, 
by comparing the undiluted background distortion
signal with the radial distortion profile of color-magnitude (or
color-color) space occupied by the cluster members,
but including inevitable background
galaxies falling in the same space\cite{2007ApJ...663..717M}.
This technique allows the light profile of the cluster to be determined in
a way which is independent of the number density fluctuations
in the background population, which otherwise limit the
calculation of the cluster light profiles and luminosity functions
from counts of member galaxies.
The resulting light profile can be compared with the mass profile 
to examine the radial behavior of
$M/L$\cite{2007ApJ...663..717M,Medezinski+2009}.

\subsection{Aperture Mass Densitometry}

The observed image distortion of background galaxies can be directly 
used to derive the projected gravitational mass of clusters.
The aperture mass estimate within the angular radius $\theta$,
 $M_{\rm \zeta}(<\theta)$,
in terms of the tangential component $\gamma_+$ of the gravitational shear
can be expressed as
\begin{equation}
M_{\zeta}(<\theta)=\pi (D_d\theta)^2 \Sigma_{\rm crit}\zeta(\theta),
\label{eq:zetamass}
\end{equation}
using the aperture-mass densitometry, or the so-called
``$\zeta$-statistic''\cite{1994ApJ...437...56F},
defined as
\begin{equation}
\zeta(\theta)\equiv \frac{2}{ 1- \theta^2/\theta_{\rm
out}^2} 
\int_{\theta}^{\theta_{\rm out}} \! d\ln\theta \, 
\langle \gamma_+(\theta) \rangle
= \overline{\kappa}(<\theta)- \overline{\kappa}(\theta,\theta_{\rm out}),
\label{eq:zetastat}
\end{equation}
where 
$\theta_{\rm out} (>\theta)$ is the (constant) outer background radius,
$\langle 
\gamma_+(\theta)\rangle=\overline{\kappa}(<\theta)-\langle\kappa(\theta)
\rangle$ is an azimuthal average of the tangential component of the
gravitational shear at radius $\theta$, 
and $\overline{\kappa} \equiv 
{\overline{\Sigma}_m}/{\Sigma_{\rm crit}}$ 
is the mean convergence.
Errors on $\zeta(\theta)$ are calculated by propagating 
the rms errors $\sigma_+(\theta_n)$ (equation [\ref{eq:err_gtx}]) 
for the tangential shear measurement.
Equations (\ref{eq:zetamass}) and (\ref{eq:zetastat})
show that the cluster mass can be measured from 
the galaxies ellipticity within the annulus bounded by $\theta$ and 
$\theta_{\rm out}$ located just {\it outside} the mass to be measured. 
In the weak lensing regime  where $|\kappa|,|\gamma|\ll 1$, $\langle
\gamma_+(\theta)\rangle$ is observable: $\langle
g_+(\theta)\rangle\approx \langle \gamma_+(\theta)\rangle$.

As revealed in equation.~(\ref{eq:zetastat}), the $\zeta-$statistic 
yields a mean convergence interior to $\theta$,
 subtracted by the mean background within the annulus 
between $\theta$ and $\theta_{\rm out}$,
$\overline{\kappa}(<\theta) - \overline{\kappa}(\theta,\theta_{\rm out})$. 
Hence, as long as
$\overline{\kappa} (\theta,\theta_{\rm out}) \ll
\overline{\kappa} (<\theta)$, the enclosed mass within $\theta$
can be obtained by multiplying $\zeta$ by 
the area $\pi \Sigma_{\rm crit} (D_d \theta)^2$.  
The inner radius $\theta$ can almost be arbitrarily chosen to obtain the
aperture mass interior to $\theta$, as long as the weak lensing
approximation is valid.
Obviously, the aperture mass  
is smaller than the enclosed mass 
by a negative compensating mass that serves 
to remove the contribution from a background uniform mass sheet, and
the degree of deviation depends on 
how steep the density profile is.

The $\zeta$-statistic is a circular aperture mass estimator.  
It gives rise to some errors for irregular and merging clusters.
Nevertheless, this effect has been estimated less than
$10\%$\cite{2004MNRAS.350.1038C}. 
The choice of the parameter $\theta_{\rm out}$ may also affect the
cluster mass estimate in a practical application.  For example, 
a small $\theta_{\rm out}$ will generate large Poisson noise 
since the galaxy number for a shear estimate within the annulus bound
by $\theta$ and $\theta_{\rm out}$ is not sufficiently large.  
On the other hand, if $\theta_{\rm out}$ is too large, 
the cluster mass measurement can be contaminated by neighboring
clusters in projection. 
For a projected lens system, $\langle \gamma_+ \rangle$ 
is produced not only by the cluster itself but also 
by the projected neighboring clusters and large scale structure.

Clowe et al.~\cite{2000ApJ...539..540C}
proposed a variant of aperture mass densitometry\cite{1994ApJ...437...56F},
by introducing two parameters to specify the
annular background region, of the form:
\begin{eqnarray}
\label{eq:zeta}
\zeta_{\rm c}(\theta)
&\equiv &
2\int_{\theta}^{\theta_{\rm inn}}\!d\ln\theta' 
\langle \gamma_+(\theta')\rangle \nonumber
+ \frac{2}{1-(\theta_{\rm inn}/\theta_{\rm out})^2}\int_{\theta_{\rm
inn}}^{\theta_{\rm out}}\! d\ln\theta'  
\langle\gamma_+(\theta')\rangle
\nonumber\\
&=& 
\overline{\kappa}(<\theta) - \overline{\kappa}
(\theta_{\rm inn},\theta_{\rm out}),
\end{eqnarray} 
where 
 $\theta_{\rm inn}$ and $\theta_{\rm out}$ 
($\theta_{\rm  out}> \theta_{\rm inn}>\theta$) are the inner and
outer radii of the annular background region in which the mean
background contribution, 
$\bar{\kappa}_b\equiv
\bar{\kappa}(\theta_{\rm inn},\theta_{\rm out})$,
is defined.
This cumulative mass estimator is often referred to as 
the $\zeta_{\rm  c}$-statistic. 
This cumulative mass estimator
subtracts  from the mean convergence 
$\bar{\kappa}( \theta)$
a constant 
$ \bar{\kappa}_{\rm b}$
for all apertures $\theta$ in the measurements, 
thus removing any DC component in the control
region $\theta = [\theta_{\rm inn}, \theta_{\rm out}]$.
We note that the $\bar{\kappa}_b$ is 
a non-observable free parameter.
The $\zeta_{\rm c}$-statistic will be particularly useful for wide-field
imaging observations, such as with the Suprime-Cam on the Subaru
telescope ($34'\times 27'$) 
and the Megacam on the CFHT ($1^\circ\times 1^\circ$), in which one can
identify a background region well outside of the cluster region.

\subsection{Weak Lensing Convergence}

Umetsu \& Broadhurst~\cite{UB2008}
have developed a non-parametric method for
reconstructing the one-dimensional $\kappa$-profile utilizing
the $\zeta_{\rm c}$-statistic measurement.
Unlike strong-lensing based boundary
conditions\cite{BTU+05},
this method utilizes an outer boundary condition on the mean background
density $\bar\kappa_b$ to derive a $\kappa$-profile (see Schneider \&
Seitz 1995 for an alternative method for a direct inversion of the mass
profile).\footnote{The mass-sheet degeneracy is inevitable in any mass
reconstruction method based solely on the shearing effect.}
For a given boundary condition $\bar{\kappa}_b$, the average
convergence $\overline{\kappa}(<\theta)$ is estimated as
$\overline{\kappa}(<\theta) = \zeta_{\rm c}(\theta) + \bar{\kappa}_b$.
Then, we define a discretized estimator for $\kappa$ as
\begin{equation}
\label{eq:zeta2kappa}
\kappa(\overline{\theta}_m)
= \alpha_2^m \zeta(\theta_{m+1})
- \alpha_1^m \zeta(\theta_{m})
+
\bar{\kappa}_b,
\end{equation}
where
\begin{equation}
\alpha_1^m = \frac{1}{2\Delta\ln\theta_m} 
\left( 
  \frac{\theta_{m}}{ \overline{\theta}_m }
\right)^2, \, \, 
\alpha_2^m = \frac{1}{2\Delta\ln\theta_m} 
\left(\frac{\theta_{m+1}}{ \overline{\theta}_m }\right)^2,
\end{equation}
and 
 $\bar\theta_m$ is the weighted center of the $m$th radial bin
bounded by $(\theta_m,\theta_{m+1})$
(see Refs.~\cite{UB2008} and \cite{Umetsu+2010_CL0024}).
The error covariance matrix $C_{mn}$ of $\kappa_m$ is expressed as
\begin{eqnarray}
C_{mn} 
=
\alpha_2^m \alpha_2^n C^{\zeta}_{m+1,n+1}
+  
\alpha_1^m \alpha_1^n C^{\zeta}_{m,n}\nonumber
-
\alpha_1^m \alpha_2^n C^{\zeta}_{m,n+1}
-
\alpha_2^m \alpha_1^n C^{\zeta}_{m+1,n},
\end{eqnarray}
where $C^{\zeta}_{mn}\equiv \langle \delta\zeta_m\delta\zeta_n\rangle$
is the bin-to-bin error covariance matrix of the aperture densitometry
measurements which is calculated by propagating the rms errors
$\sigma_+(\theta_m)$ for the tangential shear measurement (see \S
\ref{sec:gtx}).  
In the non-linear regime, however,
the $\gamma_+(\theta)$ is not a direct observable.
Therefore, non-linear corrections need to be taken into account 
in the mass reconstruction process.
In the subcritical regime (i.e., outside the critical curves),
the $\gamma_+(\theta)$
can be expressed in terms of the 
the averaged tangential reduced shear as
$\langle g_+(\theta) \rangle \approx
\gamma_+(\theta)/[1-\kappa(\theta)]$
assuming a quasi circular symmetry in the projected mass distribution.
This non-linear equation (\ref{eq:zeta})
for $\zeta_{\rm c}(\theta)$ can be solved by an iterative procedure,
which is outlined in Ref.~\cite{Umetsu+2010_CL0024}.

\begin{figure}[htb]
\begin{center}
\includegraphics[width=130mm,angle=0]{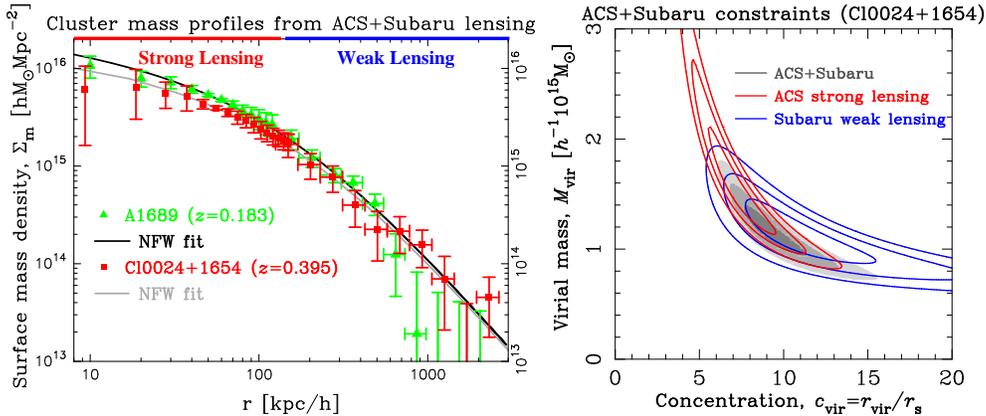}
\end{center}
\caption{
{\it Left}:
Cluster surface mass density profiles 
for A1689 ({\it green triangles}) and CL0024+1654
({\it red squares}) reconstructed from strong lensing (HST/ACS: $r\simgt
 200\,$kpc) and  weak lensing measurements (Subaru: $r\simgt
 200$\,kpc). The curvature of these profiles demonstrates that both strong
and weak lensing data are required to make an accurate
 measurement of the cluster mass structure parameters, such as the halo
 virial mass, $M_{\rm vir}$, and concentration, $c_{\rm vir}$. 
Also shown as solid curves are the best-fit NFW models,
with a continuously
 steepening density profile,
for A1689 ({\it
 black}) and CL0024+1654 ({\it gray}).
{\it Right}:
Joint constraints on the  
NFW model parameters ($c_{\rm vir},M_{\rm vir}$)
derived for CL0024+1654. 
The red and blue contours show the 68.3\%, 95.4\%, and
99.7\% confidence limits
for the inner strong lensing and outer weak lensing data,
 respectively. Combining strong and weak-lensing information 
({\it gray contours}) significantly reduces the uncertainties on the
 profile parameters. Figures taken and modified from Umetsu \&
 Broadhurst~\cite{UB2008} and Umetsu, Medezinski, Broadhurst et
 al.~\cite{Umetsu+2010_CL0024}. 
\label{fig:kprof} 
}
\end{figure}

In the left panel of 
Figure \ref{fig:kprof} we show cluster surface mass density profiles 
$\Sigma_m(\theta)$ for A1689\cite{UB2008} and
CL0024+1654\cite{Umetsu+2010_CL0024} 
as reconstructed from combined Subaru weak-lensing ($r\simgt 200$\,kpc)
and HST/ACS strong-lensing ($r\simlt 200$\,kpc)
observations.
The joint mass profiles for the clusters
continuously steepen out to their virial radii, and are well fitted with
single NFW profiles, which provide a good description of the equilibrium
density profiles of collisionless DM halos in cosmological
$N$-body simulations.
The right panel of Figure \ref{fig:kprof} shows two-dimensional
marginalized constraints on the NFW model parameters ($c_{\rm
vir},M_{\rm vir}$) derived for CL0024+1654 from the joint mass profile
shown in the left panel. This figure demonstrates 
that combining strong and weak-lensing information 
({\it grey contours}) significantly reduces the uncertainties on the
 profile parameters. 
Such non-parametric mass profiles are also useful when comparing
the total matter distribution with cluster properties obtained from
other
wavelengths and/or
approaches\cite{2008MNRAS.386.1092L,Lemze+2009,Lapi+Cavaliere2009,Peng+2009}.  

\appendix

\section{Multiple Lens Equation}
\label{appendix:multi}
 
We may discretize the cosmological lens equation
(\ref{eq:cosmo_lenseq2}) by dividing the radial integral between the
source ($\chi=\chi_s$) and the observer ($\chi=0$) into $N$-comoving
boxes
($N-1$ lens-planes)
separated by a comoving distance of $\Delta\chi$. 
The angular position $\btheta^{(n)}$
of a light ray at the $n$th plane ($n\le N$) is then given by
\begin{equation}
\btheta^{(n)}-\btheta^{(0)}=-\sum_{p=1}^{n-1} 
\frac{r(\chi_n-\chi_p)}
{r(\chi_n)}
\bnabla_{\perp} \psi_p
\end{equation}
where $\psi_p$ is the effective lensing potential of the $p$th lens
plane
($p=1,2,..,N-1$):
\begin{equation}
\psi_p = \frac{2}{c^2} \int_{\chi_p}^{\chi_p+\Delta\chi}\! d\chi\,\Psi.
\end{equation}
The Jacobian matrix is expressed as
\begin{equation}
\bA^{(n)} := \frac{\partial \btheta^{(n)}}
                  {\partial \btheta^{(0)}}
=
\bI-\sum_{p=1}^{n-1} g(\chi_p,\chi_n) {\bH}^{(p)} {\bA}^{(p)}
\equiv \bI-\bPsi^{(n)}
\end{equation}   
where $H_{\alpha\beta}^{(p)}=\partial^2 \psi_p/\partial\chi^{\perp,\alpha}
\partial\chi^{\perp,\beta}$ ($\alpha,\beta=1,2$) is the
Hessian matrix, and $g(\chi_p,\chi_n)$ is the effective lensing distance
for the $p$th lens plane.
In general
(without the Boron approximation), 
the $2\times 2$ Jacobian matrix can be decomposed into
the following form:
\begin{equation}
{\cal A}_{\alpha\beta}=\delta_{\alpha\beta}-\psi_{\alpha\beta}=
(1-\kappa)\delta_{\alpha\beta} - \gamma_1\sigma_3 - \gamma_2\sigma_1 -i
 \omega \sigma_2
\end{equation}
where 
$\kappa = (\psi_{11}+\psi_{22})/2$,
$\gamma_1=(\psi_{11}-\psi_{22})/2$,
$\gamma_2=(\psi_{12}+\psi_{21})/2$,
and $\omega$ is the net rotation $\omega=(\psi_{12}-\psi_{21})/2$.
The Born approximation $\bA^{(p)}=\bI$ leads to the symmetric Jacobian
matrix.


\acknowledgments
I am very grateful to
the organizers of the Fermi summer school, 
Y. Rephaeli and
A. Cavaliere,
for the invitation to 
present the research discussed in this review.
I would like to thank Masahiro Takada and Patrick Koch for 
providing invaluable comments on the manuscript.
This work is partially supported by
the National Science Council of Taiwan
under the grant NSC97-2112-M-001-020-MY3.


\begin{thebibliography}{0}
\bibitem{1990ApJ...349L...1T}
\BY{Tyson J.~A., Wenk R.~A. \atque Valdes F.}
\IN{\apjl}{349}{1990}{L1}.

\bibitem{1993ApJ...404..441K}
\BY{Kaiser N. \atque Squires G.}
\IN{\apj}{404}{1993}{441}.

\bibitem{1995A&A...294..411S}
\BY{Schneider P. \atque Seitz C.}
\IN{\aap}{294}{1995}{411}.

\bibitem{2001PhR...340..291B}
\BY{Bartelmann M. \atque Schneider P.}
\IN{\physrep}{340}{2001}{291}.

\bibitem{1999PThPS.133...53U}
\BY{Umetsu K., Tada M. \atque Futamase T.}
\IN{Progress of Theoretical Physics Supplement}{133}{1999}{53}{ (arXiv:astro-ph/0004400)}.

\bibitem{Clowe+2006_Bullet}
\BY{Clowe D., Brada{\v c} M., Gonzalez A.~H., Markevitch M.,
  Randall S.~W., Jones C. \atque Zaritsky D.}
  \IN{\apjl}{648}{2006}{L109}. 

\bibitem{Okabe&Umetsu08}
\BY{Okabe N. \atque Umetsu K.}
\IN{\pasj}{60}{2008}{345}.

\bibitem{BTU+05}
\BY{Broadhurst T., Takada M., Umetsu K., Kong X., Arimoto N.,
  Chiba M. \atque Futamase T.}
  \IN{\apjl}{619}{2005}{L143}.

\bibitem{UB2008}
\BY{Umetsu K. \atque Broadhurst T.}
\IN{\apj}{684}{2008}{177}{ (arXiv:0712.3441)}. 

\bibitem{2007ApJ...668..643L}
\BY{Limousin M., Richard J., Jullo E., Kneib J.-P., Fort B.,
  Soucail G., El{\'{\i}}asd{\'o}ttir {\'A}., Natarajan P., Ellis R.~S.,
  Smail I., Czoske O., Smith G.~P., Hudelot P., Bardeau S., Ebeling
  H., Egami E. \atque Knudsen K.~K.}
  \IN{\apj}{668}{2007}{643}.
 
\bibitem{2008arXiv0805.2552M}
\BY{Mandelbaum R., Seljak U. \atque Hirata C.~M.}
\IN{\jcap}{8}{2008}{6}{ (arXiv:0805.2552)}. 

\bibitem{BUM+08}
\BY{Broadhurst T., Umetsu K., Medezinski E., Oguri M. \atque
  Rephaeli Y.}
  \IN{\apjl}{685}{2008}{L9}. 

\bibitem{2009arXiv0903.1103O}
\BY{Okabe N., Takada M., Umetsu K., Futamase T. \atque Smith
  G.~P.}
  \IN{\pasj, submitted}{}{2009}{ arXiv:0903.1103}. 

\bibitem{Oguri+2009_Subaru}
\BY{Oguri M., Hennawi J.~F., Gladders M.~D., Dahle H., Natarajan
  P., Dalal N., Koester B.~P., Sharon K. \atque Bayliss M.}
  \IN{\apj}{699}{2009}{1038}.

\bibitem{2007arXiv0709.1159J}
\BY{Johnston D.~E., Sheldon E.~S., Wechsler R.~H., Rozo E., Koester
  B.~P., Frieman J.~A., McKay T.~A., Evrard A.~E., Becker M.~R. \atque
  Annis J.} {(2007)} {arXiv:0709.1159}

\bibitem{2007ApJS..170..377S}
\BY{Spergel D.~N., Bean R., Dor{\'e} O., Nolta M.~R., Bennett
  C.~L., Dunkley J., Hinshaw G., Jarosik N., Komatsu E., Page L.,
  Peiris H.~V., Verde L., Halpern M., Hill R.~S., Kogut A., Limon
  M., Meyer S.~S., Odegard N., Tucker G.~S., Weiland J.~L., Wollack
  E. \atque Wright E.~L.}
  \IN{\apjs}{170}{2007}{377}.

\bibitem{2008arXiv0803.0547K}
\BY{Komatsu E., Dunkley J., Nolta M.~R., Bennett C.~L., Gold B.,
  Hinshaw G., Jarosik N., Larson D., Limon M., Page L., Spergel
  D.~N., Halpern M., Hill R.~S., Kogut A., Meyer S.~S., Tucker G.~S.,
  Weiland J.~L., Wollack E. \atque Wright E.~L.}
  \IN{\apjs}{180}{2009}{330}.

\bibitem{1996MNRAS.283..837S}
\BY{Schneider P.}
\IN{\mnras}{283}{1996}{837}.

\bibitem{2000A&A...355...23E}
\BY{Erben T., van Waerbeke L., Mellier Y., Schneider P., Cuillandre
  J.-C., Castander F.~J. \atque Dantel-Fort M.}
  \IN{\aap}{355}{2000}{23}.  

\bibitem{2000ApJ...539L...5U}
\BY{Umetsu K. \atque Futamase T.}
\IN{\apjl}{539}{2000}{L5}{ (arXiv:astro-ph/0004373)}.  

\bibitem{2002ApJ...580L..97M}
\BY{Miyazaki S., Hamana T., Shimasaku K., Furusawa H., Doi M.,
  Hamabe M., Imi K., Kimura M., Komiyama Y., Nakata F., Okada N.,
  Okamura S., Ouchi M., Sekiguchi M., Yagi M. \atque Yasuda N.}
  \IN{\apjl}{580}{2002}{L97}. 

\bibitem{2007ApJ...669..714M}
\BY{Miyazaki S., Hamana T., Ellis R.~S., Kashikawa N., Massey
  R.~J., Taylor J. \atque Refregier A.}
  \IN{\apj}{669}{2007}{714}. 

\bibitem{Hamana+2009}
\BY{Hamana T., Miyazaki S., Kashikawa N., Ellis R.~S., Massey
  R.~J., Refregier A. \atque Taylor J.~E.}
  \IN{\pasj}{61}{2009}{833}. 

\bibitem{SEF1992}
\BY{Schneider P., Ehlers J. \atque Falco E.~E.}
\IN{Gravitational Lenses (Berlin: Springer, Verlag)}{}{1992}{}. 

\bibitem{1992ARA&A..30..311B}
\BY{Blandford R.~D. \atque Narayan R.}
\IN{\araa}{30}{1992}{311}. 

\bibitem{refsdal_surdej}
\BY{Refsdal S. \atque Surdej J.}
\IN{Rep.~Prog.~Phys.}{57}{1994}{117}. 

\bibitem{1996astro.ph..6001N}
\BY{Narayan R. \atque Bartelmann M.}
\IN{arXiv:astro-ph/9606001}{}{1996}{}.

\bibitem{1999PThPS.133....1H}
\BY{Hattori M., Kneib J. \atque Makino N.}
\IN{Progress of Theoretical Physics Supplement}{133}{1999}{1}.  

\bibitem{1995PhRvL..75.1439L}
\BY{Lebach D.~E., Corey B.~E., Shapiro I.~I., Ratner M.~I., Webber
  J.~C., Rogers A.~E.~E., Davis J.~L. \atque Herring T.~A.}
  \IN{Physical Review Letters}{75}{1995}{1439}.

\bibitem{1985A&A...143..413S}
\BY{Schneider P.}
\IN{\aap}{143}{1985}{413}.

\bibitem{1993PThPh..90..753S}
\BY{Sasaki M.}
\IN{Progress of Theoretical Physics}{90}{1993}{753}.

\bibitem{1994CQGra..11.2345S}
\BY{Seitz S., Schneider P. \atque Ehlers J.}
\IN{Classical and Quantum Gravity}{11}{1994}{2345}.

\bibitem{1995PThPh..93..647F}
\BY{Futamase T.}
\IN{Progress of Theoretical Physics}{93}{1995}{647}.

\bibitem{takada_phd}
\BY{Takada M.}
\IN{Ph.D. thesis, Tohoku University, Japan}{}{2000}{}. 

\bibitem{2003MNRAS.339.1155H}
\BY{Hoekstra H.}
\IN{\mnras}{339}{2003}{1155}.

\bibitem{Keeton2001}
\BY{Keeton C.~R.}
\IN{arXiv:astro-ph/0102341}{}{2001}.

\bibitem{1997ApJ...490..493N}
\BY{Navarro J.~F., Frenk C.~S. \atque White S.~D.~M.}
\IN{\apj}{490}{1997}{493}.

\bibitem{1996A&A...313..697B}
\BY{Bartelmann M.}
\IN{\aap}{313}{1996}{697}. 

\bibitem{2000ApJ...534...34W}
\BY{Wright C.~O. \atque Brainerd T.~G.}
\IN{\apj}{534}{2000}{34}. 

\bibitem{Umetsu+2010_CL0024}
\BY{Umetsu K., Medezinski E., Broadhurst T., Zitrin A., Okabe N.,
  Hsieh B. \atque Molnar S.~M.}
  \IN{\apj, in press}{}{2010}{arXiv:0908.0069}. 

\bibitem{2007MNRAS.374.1377T}
\BY{Taylor A.~N., Kitching T.~D., Bacon D.~J. \atque Heavens A.~F.}
\IN{\mnras}{374}{2007}{1377}. 

\bibitem{2007ApJ...663..717M}
\BY{Medezinski E., Broadhurst T., Umetsu K., Coe D., Ben{\'{\i}}tez
  N., Ford H., Rephaeli Y., Arimoto N. \atque Kong X.}
  \IN{\apj}{663}{2007}{717}.

\bibitem{2002ApJ...568...20C}
\BY{Crittenden R.~G., Natarajan P., Pen U.-L. \atque Theuns T.}
\IN{\apj}{568}{2002}{20}. 

\bibitem{2006MNRAS.365..414B}
\BY{Bacon D.~J., Goldberg D.~M., Rowe B.~T.~P. \atque Taylor A.~N.}
\IN{\mnras}{365}{2006}{414}. 

\bibitem{2007ApJ...660..995O}
\BY{Okura Y., Umetsu K. \atque Futamase T.}
\IN{\apj}{660}{2007}{995}{ (arXiv:astro-ph/0607288)}. 

\bibitem{2008ApJ...680....1O}
\BY{Okura Y., Umetsu K. \atque Futamase T.}
\IN{\apj}{680}{2008}{1}{ (arXiv:0710.2262)}. 

\bibitem{1995ApJ...439L...1K}
\BY{Kaiser N.}
\IN{\apjl}{439}{1995}{L1}.

\bibitem{2000ApJ...530..547J}
\BY{Jain B., Seljak U. \atque White S.}
\IN{\apj}{530}{2000}{547}. 

\bibitem{2002PASJ...54..833M}
\BY{Miyazaki S., Komiyama Y., Sekiguchi M., Okamura S., Doi M.,
  Furusawa H., Hamabe M., Imi K., Kimura M., Nakata F., Okada N.,
  Ouchi M., Shimasaku K., Yagi M. \atque Yasuda N.}
  \IN{\pasj}{54}{2002}{833}.

\bibitem{1995ApJ...449..460K}
\BY{Kaiser N., Squires G. \atque Broadhurst T.}
\IN{\apj}{449}{1995}{460}.

\bibitem{2001A&A...366..717E}
\BY{Erben T., Van Waerbeke L., Bertin E., Mellier Y. \atque
  Schneider P.}
  \IN{\aap}{366}{2001}{717}.

\bibitem{2006MNRAS.368.1323H}
\BY{Heymans C., Van Waerbeke L., Bacon D., Berge J., Bernstein G.,
  Bertin E., Bridle S., Brown M.~L., Clowe D., Dahle H., Erben T.,
  Gray M., Hetterscheidt M., Hoekstra H., Hudelot P., Jarvis M.,
  Kuijken K., Margoniner V., Massey R., Mellier Y., Nakajima R.,
  Refregier A., Rhodes J., Schrabback T. \atque Wittman D.}
  \IN{\mnras}{368}{2006}{1323}. 

\bibitem{2007MNRAS.376...13M}
\BY{Massey R., Heymans C., Berg{\'e} J., Bernstein G., Bridle S.,
  Clowe D., Dahle H., Ellis R., Erben T., Hetterscheidt M., High
  F.~W., Hirata C., Hoekstra H., Hudelot P., Jarvis M., Johnston D.,
  Kuijken K., Margoniner V., Mandelbaum R., Mellier Y., Nakajima R.,
  Paulin-Henriksson S., Peeples M., Roat C., Refregier A., Rhodes J.,
  Schrabback T., Schirmer M., Seljak U., Semboloni E. \atque van
  Waerbeke L.}
  \IN{\mnras}{376}{2007}{13}.

\bibitem{1995ApJ...438...49B}
\BY{Broadhurst T.~J., Taylor A.~N. \atque Peacock J.~A.}
\IN{\apj}{438}{1995}{49}. 

\bibitem{1998ApJ...501..539T}
\BY{Taylor A.~N., Dye S., Broadhurst T.~J., Benitez N. \atque van
  Kampen E.}
  \IN{\apj}{501}{1998}{539}.

\bibitem{1995ApJ...446L..55T}
\BY{Tyson J.~A. \atque Fischer P.}
\IN{\apjl}{446}{1995}{L55+}.

\bibitem{2003ApJ...597...98H}
\BY{Hamana T., Miyazaki S., Shimasaku K., Furusawa H., Doi M.,
  Hamabe M., Imi K., Kimura M., Komiyama Y., Nakata F., Okada N.,
  Okamura S., Ouchi M., Sekiguchi M., Yagi M. \atque Yasuda N.}
  \IN{\apj}{597}{2003}{98}.

\bibitem{Okabe+Umetsu2008}
\BY{Okabe N. \atque Umetsu K.}
\IN{\pasj}{60}{2008}{345}.

\bibitem{2005ApJ...619L.143B}
\BY{Broadhurst T., Takada M., Umetsu K., Kong X., Arimoto N.,
  Chiba M. \atque Futamase T.}
  \IN{\apjl}{619}{2005}{L143}.

\bibitem{2009ApJ...694.1643U}
\BY{Umetsu K., Birkinshaw M., Liu G.-C., Wu J.-H.~P., Medezinski
  E., Broadhurst T., Lemze D., Zitrin A., Ho P.~T.~P., Huang
  C.-W.~L., Koch P.~M., Liao Y.-W., Lin K.-Y., Molnar S.~M., Nishioka
  H., Wang F.-C., Altamirano P., Chang C.-H., Chang S.-H., Chang
  S.-W., Chen M.-T., Han C.-C., Huang Y.-D., Hwang Y.-J., Jiang H.,
  Kesteven M., Kubo D.~Y., Li C.-T., Martin-Cocher P., Oshiro P.,
  Raffin P., Wei T. \atque Wilson W.}
  \IN{\apj}{694}{2009}{1643}{ (arXiv:0810.0969)}.

\bibitem{2005PhRvL..95x1302Z}
\BY{Zhang P. \atque Pen U.}
\IN{Physical Review Letters}{95}{2005}{241302}. 

\bibitem{Medezinski+2009}
\BY{Medezinski E., Broadhurst T., Umetsu K., Oguri M., Rephaeli Y.
  \atque Ben{\'{\i}}tez N.}
  \IN{\apj, in press}{}{2010}{arXiv:0906.4791}.

\bibitem{2004MNRAS.350.1038C}
\BY{Clowe D., De Lucia G. \atque King L.}
\IN{\mnras}{350}{2004}{1038}. 

\bibitem{2000ApJ...539..540C}
\BY{Clowe D., Luppino G.~A., Kaiser N. \atque Gioia I.~M.}
\IN{\apj}{539}{2000}{540}. 

\bibitem{1994ApJ...437...56F}
\BY{Fahlman G., Kaiser N., Squires G. \atque Woods D.}
\IN{\apj}{437}{1994}{56}. 

\bibitem{2008MNRAS.386.1092L}
\BY{Lemze D., Barkana R., Broadhurst T.~J. \atque Rephaeli Y.}
\IN{\mnras}{386}{2008}{1092}.

\bibitem{Lemze+2009}
\BY{Lemze D., Broadhurst T., Rephaeli Y., Barkana R. \atque Umetsu
  K.}
  \IN{\apj}{701}{2009}{1336}.

\bibitem{Lapi+Cavaliere2009}
\BY{Lapi A. \atque Cavaliere A.}
\IN{\apjl}{695}{2009}{L125}.

\bibitem{Peng+2009}
\BY{Peng E., Andersson K., Bautz M.~W. \atque Garmire G.~P.}
\IN{\apj}{701}{2009}{1283}.
\end{thebibliography}
\end{document}